\documentclass[debug]{rmaa}

\usepackage{graphicx}
\usepackage{subfigure}
\usepackage{hyperref}
\usepackage{paralist}

\usepackage{psfrag,color}

\usepackage[latin1]{inputenc}

\title{Detecting the growth of structures  in Pure Stellar Disk Models} 

\author{
  D. Valencia-Enr\'\i quez,\altaffilmark{1} 
  I. Puerari,\altaffilmark{1}
  L. Chaves-Velasquez\altaffilmark{1}}
  
\altaffiltext{1}{Instituto Nacional de Astrof\'{i}sica, \'{O}ptica y Electr\'{o}nica, INAOE, M\'exico.}

\shortauthor{Valencia-Enr\'\i quez, Puerari, \& Chaves-Velasquez}
\shorttitle{Detecting the growth of structures}

\fulladdresses{
\item Diego Valencia-Enr\'\i quez, Iv\^anio Puerari and Leonardo Chaves-Velasquez: Instituto Nacional de Astrof\'{i}sica, 
      \'{O}ptica y Electr\'{o}nica, Calle Luis Enrique Erro 1, 
      Santa Mar\'{i}a Tonantzintla, 72840 Puebla, Mexico
      (E-mail: valencia@inaoep.mx, diegovalencia5@gmail.com).}

\listofauthors{D. Valencia-Enr\'\i quez, I. Puerari, \& L. Chaves-Velasquez}
\indexauthor{Valencia-Enr\'\i quez, D.}
\indexauthor{Puerari, I.}
\indexauthor{Chaves-Velasquez, L.}

\abstract{We performed a series of 3D N-body simulations
          where the initial conditions  were chosen  to get
          two sets of models; unbarred and barred ones.
          In this work, we analyze the growth of spirals and
          bar structures using 1D, and 2D Fourier Transforms FT methods. 
          Spectrograms and diagrams of the amplitude of the Fourier 
          coefficients as a function of time, 
          radius and pitch angle show that the general morphology, 
          of our modeled galaxies, is due to the superposition of
          structures which have different values of pitch angle 
          and number of arms. 
          Also, we made in barred models a geometric classification of 
          orbits from the bar reference frame showing 
          that the barred potential and the Lagrangian points 
          $L_4$ and $L_5$ catch approximately one-third of the total disk mass. 
          
	}

\resumen{Se ejecutaron simulaciones de N-cuerpos en 3D 
	donde las condiciones iniciales fueron escogidas 
	para obtener dos conjuntos de modelos;
	no-barrados y barrados. Aqu\'i, Se analiza el 
	crecimiento de las estructuras espirales y/o barra
	usando m\'etodos de Transformada de Fourier 1D y 2D.  
	Los espectrogramas y los diagramas de la amplitud de 
	los coeficientes de Fourier en funci\'on del tiempo, 
	radio y \'angulo de enrollamiento muestra que
	la morfolog\'\i a de nuestros modelos 
	es devida a la superposici\'on de
	diversas estrucuturas con valores diferentes 
	para el \'angulo de enrollamiento,
	n\'umero de brazos y velocidad angular. 
	Adem\'as, en los modelos donde se forma una barra,
	se estudi\'o las \'orbitas de las part\'\i 
	culas en el sistema de referencia de esta.
	Una clasificaci\'on geom\'etrica de las \'orbitas 
	muestra que el potencial de la barra y los 
	puntos Lagrangianos $L_4$ y $L_5$ 
	capturan aproximadamente 1/3 de la masa total del disco.
	}

\addkeyword{galaxies: kinematics and dynamics}
\addkeyword{galaxies: spiral}
\addkeyword{galaxies: structure}
\addkeyword{methods: numerical}

\begin{document}
% Typeset article header
\maketitle

\section{Introduction}
\label{introduction}

The physical origin and the evolution of non-axisymmetric 
structures in disk galaxies are long-standing problems 
in Astrophysics. One of the most widely accepted 
hypothesis is the Spiral Density Wave Theory  
(e.g. \citealt{1964ApJ...140..646L, 1996ssgd.book.....B}). 
In this theory, the spiral arms are explained 
as long-lived quasi-stationary 
density waves with a constant pattern speed. 
Additionally, \citet{1996ssgd.book.....B} introduced the
supposition that these waves are the result of global 
modes. Their global mode analysis shows that the spiral arms 
are manifestations of the gravitationally unstable global 
eigen-oscillations of 
the disk galaxies (\citealt{dobbsbaba2014}).
\citet{1965MNRAS.130..125G}, \citet{1966ApJ...146..810J} and
\citet{1981seng.proc..111T} 
proposed that the spiral arms be 
stochastically produced by local gravitational 
perturbations in a differentially rotating disk: 
short leading spiral perturbation shears at corotation 
into a short trailing spiral due to the differential rotation. 
The wave is amplified by self-gravity of the assembly of stars
at the perturbation.  This mechanism is known as 
swing-amplification, and
the resulting spiral structure is short lived 
(\citealt{1981seng.proc..111T}).
\citet{1997A&A...322..442M} 
suggested that global modes in stellar disks 
can be coupled through non-linear interactions.
They proposed that a wave 1 excites a wave 2 
through second-order coupling terms that are 
large when CR of a wave 1 lies at approximately 
the same radius as the ILR of wave 2
(\citealt{2013pss5.book..923S}).

Many simulations of stellar disks 
show that the spiral arms fade out 
after some galactic rotations. 
Furthermore, if the effects of gas
are not included, the velocity 
dispersion of the disk will increase; 
therefore the disk will become stable 
and will not form spiral arms
(e.g. \citealt{1984ApJ...282...61S, 
2009ApJ...706..471B, 2011ApJ...735....1W}). 
\citet{1984ApJ...282...61S} noticed that 
the spiral pattern in N-body simulations 
generally fades over time because 
the spiral arm structure heats out the
disk kinematically and causes Toomre 
$Q$ parameter to rise. Hence 
the disk becomes quite stable against 
the development of non-axisymmetric 
structures. It is also shown that 
the increment of new particles with 
low-velocity dispersion at a constant 
rate in the disk maintains the spiral 
patterns for longer time scales. \citet{2009ApJ...706..471B} 
performed self-consistent high-resolution, 
N-body$+$hydrodynamical simulations to explore how 
the spiral arms are formed and maintained.
They also showed that spiral arms are not quasi-stationary, 
but they are  transient and recurrent
 like alternative theories of spiral structures suggest 
 (e.g. \citealt{1965MNRAS.130..125G, 1966ApJ...146..810J, 1981seng.proc..111T}).

\citet{2011ApJ...730..109F} performed a series of high-resolution 
3D N-body simulations of pure stellar disks. Their models
are based on those of \citet{2009ApJ...706..471B} 
being the Toomre's $Q$ initial parameter
approximately one. They showed that stellar
disks can maintain spiral features for several 
tens of rotations without the help of cooling. 
They also found that if the number of particles 
is sufficiently huge, e.g., larger than $3 \times 10^6$, 
multi-arm spirals will develop on an isolated disk and 
they can survive for more than 10 Gigayears.

\citet{2013ApJ...763...46B} discussed the growth 
of spirals structures using one very
high-resolution N-body simulation ($3\times 10^8$ particles).
They pointed out that radial migration of stars around spiral arms
are essential for damping of spiral structure because excessive Coriolis forces 
dominate the gravitational perturbation exerted by the spiral and, as a result,
stars escape from the spirals and join a new spiral at a different position. 
This process is cyclic; therefore, the dominant spiral mode indeed change 
over radius and time.

Recent works made by 
\citet{2011ApJ...735....1W}, \citet{2012MNRAS.421.1529G}, 
\citet{2013MNRAS.432.2878R}, \citet{dobbsbaba2014}
showed that the pattern speed of the spiral arms 
decreases with radius similarly to the angular 
rotation velocity of the disc. Thus, the spiral 
arms are considered to be corotating with 
the rest of the disc at every radius; 
they are material spiral arms. In these 
models, the evolution of the spiral arms 
is governed by the winding of the arms, which 
leads to breaks and bifurcations of the spiral structure.

Other works have shown that the continuous infall of substructures from the 
dark matter halos of the galaxies could induce spiral patterns by
generating localized disturbance that grows by swing amplification 
\citep{2006ApJ...653.1180G}. However, \citet{2010ApJ...709.1138D}
propose that dark matter substructures orbiting in the inner regions of the
galaxies halos would be destroyed by dynamical processes such as disk shocking,
moreover, hence would not be able to seed the formation of spiral structures.
On the other hand, the interaction with
galactic satellites could produce the growth of
the spirals (\citet{1990A&A...230...37G}, and references therein).

\citet{2013ApJ...766...34D} developed high-resolution N-body simulations to 
follow the motions of stars. Firstly, they performed the simulation by 
using equal masses for each particle of the disk; then they added
particles which have a similar mass of those of the molecular clouds. 
They demonstrated that eventually, the response of the disk can be highly 
non-linear and time variable. Ragged spiral structures can thus survive at least, in
a statistical sense, long after the original perturbing influence has been removed.

Observational evidence in spiral galaxies 
supports both long and short-lived spiral patterns. 
Recently, based on the analysis with the 
Radial Tremaine $-$ Weinberg method \citep{tremaine1984} 
using $CO$ and $H_I$ data of several galaxies, it has been  
proposed that the spiral pattern speed $\Omega_p$ may
increase with decreasing radius in some objects 
(\citealt{2006MNRAS.366L..17M, 2009ApJ...702..277M, 2012ApJ...752...52S}).
This behavior is also seeing in simulations  performed by
\citet{2011ApJ...735....1W}, \citet{2012MNRAS.421.1529G}, 
\citet{2012MNRAS.426..167G}, \citet{2013MNRAS.432.2878R}.
If this is indeed the case, the lifetime 
of the spiral structure is correspondingly very short. 
However, \citet{2013ApJ...765..105M} use azimuthal
age/color gradients across spiral arms to show 
that the spiral patterns in some grand design 
galaxies are long lived. \citet{2011ApJ...735..101F} 
estimate that the torque produced by spiral patterns
may redistribute the disk angular momentum 
in a time scale of approximately 4 Gigayears.

Recently, \citet{sahaelmegreen2016} presented 
a series of simulations
in which they changed the mass of the bulge. 
The models include the bulge
which is a King model, an exponential disk, 
and a flattened, cored dark-matter halo.
In some model with intermediate bulge mass, 
spirals structure have survived for
several Gigayears. In their models, a ``Q'' 
barrier developed in those simulations
are enough to avoid that the waves arrive at the ILR, 
and then, the wave can last for a long time.

In this work, we have generated a series of
high-resolution N-body simulations ($\sim10^6$
particles) in which we included halo, bulge,
and disk components following the distribution
functions described by \citet{1995MNRAS.277.1341K}.
The simulations were analyzed using 1D and 2D Fourier
Transform methods. These analyses show the growth
and evolution of spiral or bar structures.
This paper is organized as follow. In Section
\ref{methodology}, we describe the models and the
FT1D and FT2D methods which are used to 
illustrate the growth of non-axisymmetric
structures. In Section \ref{results}, we present
the results of our analysis, the comparison with
previous studies and a discussion. Finally, we summarize
our findings in Section \ref{summary}.

\section{Methodology}
\label{methodology}

\begin{figure}
  \centering
     \includegraphics[width=\columnwidth]{tabla_modelos}
     \caption{This Figure shows a grid of all our 26 simulations. The names of the models
	     are at the bottom left corner. The models have different disk 
             central radial velocity dispersions $\sigma_{R,0}$ and disk scale
	     height $z_d$. These values are given in the upper right corner
	     of each panel. The other parameters for the models are given in Table \ref{models_tab01}.
	     The 16 black boxes correspond to models runned  1.2 million 
             particles. The blue box represents the MW-A model \citep{1995MNRAS.277.1341K}.
             The red box shows the models runned also with 8 million particles. The model
	     in green box was also runned with different number of particles and different
             disk mass (see table 1).}
    \label{modelstab02}
\end{figure}

\begin{figure}
    \centering
    \includegraphics[width=\columnwidth]{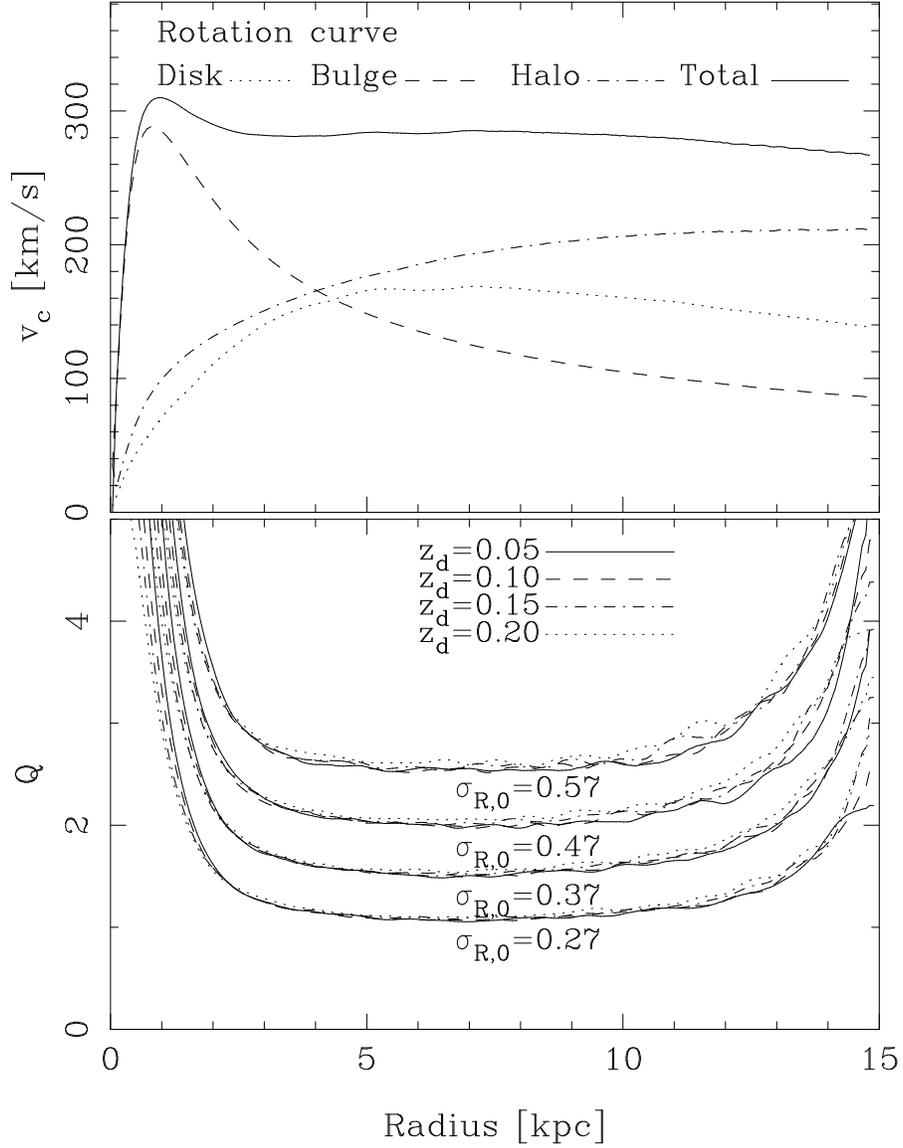}
    \caption{Upper panel: the rotation curve generated by model s27\_z10D. All our models
             have similar rotation curves, except the more massive disk one (barred models). Bottom
             panel: initial $Q$ value for the 16 unbarred models with 1.2 million particles.}
    \label{curverotation}
\end{figure}

\subsection{Setting up of the initial conditions}

We used the methodology delineated by 
\citet{1995MNRAS.277.1341K}
to generate the initial conditions of our models. In that work, they
described methods for setting up self-consistent disc-bulge-halo galaxy models. 
Our models have been evolved from 0 to 5 Gigayears, with four free parameters: 
the disk radial velocity dispersion $\sigma_R$, the disk scale 
height $z_d$, the disk mass $m_d$, and the number of particles. 
Most of the structural parameters are given in table \ref{models_tab01}.

Figure \ref{modelstab02} shows all the models
that were evolved in this work.
For all quantities, the reader can be referred to a normalization
$UL=3\ kpc$, $UT=10^7\ years$, $UV=293\ km/sec$, $UM=6\times 10^{10}\ M_\odot$,
for $G=1$.
We have two sets of simulations: unbarred and barred models.
The disk in unbarred models is stable against bar formation, 
but transient spiral structures can appear. On the other hand,
the disk in barred models is unstable to bar formation, and
it develops around the first gigayear in our simulations.
The number of particles that we use in our 
models was of  $N=1.2\times 10^6$ and $N=8\times 10^6$ particles.
The higher resolution models are made in order 
to test the effect of interactions between 
disk particles with bulge and halo particles
showing that the two-body relaxation will 
not artificially induce chaotic orbits,
which could scatter particles out of resonant orbits. Also, we run some simulations
in which the mass of the particles is the same for all the components.
We resume these parameters in Table \ref{models_tab03}.

All our models generate similar rotation curves except for those
models have the highest mass in the disk. As an example, Figure 
\ref{curverotation} shows the rotation curve resulted by model 
s27\_z10D, and the initial Toomre stability
parameter $Q$ as a function of radius generated by the first 16 models
listed on Table \ref{models_tab03}.
The models with low values of $Q$ will be called cold models. 
On the other hand, the models with high values of $Q$ will be called
hot models. We will discuss the evolution of the $Q$ parameter in section \ref{toomre}.

 \newcommand{\mc}[3]{\multicolumn{#1}{#2}{#3}}
\begin{table*}
 \begin{center}
 	\setlength{\tabnotewidth}{\columnwidth}
 	\tablecols{16}
 		% Stretch the space between table columns 
 		\setlength{\tabcolsep}{0.4\tabcolsep}
 \caption{Model parameters for the MW-A model. }
\begin{tabular}{llllllllllllllll}%\cline{2-2}\cline{4-4}
\hline \hline 
\mc{6}{l}{Disk\tabnotemark{a}} & \mc{4}{l}{Bulge\tabnotemark{b}} & \mc{6}{l}{Halo\tabnotemark{c}}\\\hline
$M_d$ & $R_d$ & $R_t$ & $z_d$ & $\delta R_{out}$ & $\sigma_{R,0}$ & 
$M_b$ &  $\Psi_c$ & $\sigma_b$ & $\rho _b$ & 
$M_h$ & $\Psi_0$ & $\sigma_0$ & $q$ & $C$ &$R_a$\\
%{\scriptsize (1)}   & {\scriptsize (2)} & {\scriptsize (3)} & {\scriptsize (4)}  & {\scriptsize (5)} & {\scriptsize (6)}& {\scriptsize (7)} & {\scriptsize (8)} & {\scriptsize (9)} & {\scriptsize (10)} &{\scriptsize (11)} & {\scriptsize (12)} & {\scriptsize (13)} &{\scriptsize (14)} &{\scriptsize (15)} &{\scriptsize (16)} \\\hline
\footnotesize{0.87}  & \footnotesize{1.0}   & \footnotesize{5.0}   & \footnotesize{0.10}  & \footnotesize{0.5}  & \footnotesize{0.47}  & 
\footnotesize{0.42}  & \footnotesize{ -2.3} & \footnotesize{0.71} & \footnotesize{14.5}  &  \footnotesize{5.2}  &  \footnotesize{-4.6}  &
\footnotesize{1.00}  & \footnotesize{1.0}   & \footnotesize{ 0.1}  & \footnotesize{0.8} \\\hline
\tabnotetext{a}{\scriptsize disk mass $M_d$,  disk scale radius $R_d$, disk truncation radius $R_t$, 
 $z_d$  disk scale height, $\delta R_{out}$ disk truncation width,  disk central radial velocity dispersion, $\sigma_{R,0}$.	}
\tabnotetext{b}{\scriptsize bulge mass $M_b$, bulge cutoff potential $\Psi_c$,  bulge velocity dispersion  $\sigma_b$,  bulge central density $\rho _b$.}
\tabnotetext{c}{\scriptsize halo mass $M_h$, halo central potential $\Psi_0$, halo velocity dispersion $\sigma_0$,  halo potential 
	flattening $q$, halo concentration $C=R_c^2/R_k^2$,  characteristic halo radius $R_a$ \citep{1994MNRAS.269...13K}.}
\end{tabular}
 \label{models_tab01}
\end{center}
\end{table*}

\begin{table*}
 \begin{center}
 	 	\setlength{\tabnotewidth}{\columnwidth}
 	 	\tablecols{11}
 	 	% Stretch the space between table columns 
 	 	\setlength{\tabcolsep}{0.4\tabcolsep}
 \caption{The table shows some parameters of the models\tabnotemark{a}. }
% \footnotesize{	
% 	\scriptsize{
\begin{tabular}{ccccccccccc}
\hline 
  model    &    $N_D$ &   $N_B$ &  $N_H$ & $N_G$    &  $m_B/m_D$ &  $m_H/m_D$ &  $M_D/M_G$ &  $M_B/M_G$ &  $M_H/M_G$ &    $M_G$  \\
\hline
s27\_z05D   &  320000 &   80000 &  800000&  1200000 &   1.92 &   2.32 &   0.14 &   0.07 &   0.80 &   6.24 \\
s27\_z10D   &  320000 &   80000 &  800000&  1200000 &   1.95 &   2.26 &   0.14 &   0.07 &   0.79 &   6.25 \\
s27\_z15D   &  320000 &   80000 &  800000&  1200000 &   2.00 &   2.25 &   0.14 &   0.07 &   0.79 &   6.24 \\
s27\_z20D   &  320000 &   80000 &  800000&  1200000 &   2.05 &   2.27 &   0.14 &   0.07 &   0.79 &   6.25 \\
s37\_z05D   &  320000 &   80000 &  800000&  1200000 &   1.92 &   2.32 &   0.14 &   0.07 &   0.80 &   6.24 \\
s37\_z10D   &  320000 &   80000 &  800000&  1200000 &   1.95 &   2.26 &   0.14 &   0.07 &   0.79 &   6.25 \\
s37\_z15D   &  320000 &   80000 &  800000&  1200000 &   2.00 &   2.25 &   0.14 &   0.07 &   0.79 &   6.24 \\
s37\_z20D   &  320000 &   80000 &  800000&  1200000 &   2.05 &   2.27 &   0.14 &   0.07 &   0.79 &   6.25 \\
s47\_z05D   &  320000 &   80000 &  800000&  1200000 &   1.92 &   2.32 &   0.14 &   0.07 &   0.80 &   6.24 \\
s47\_z10D   &  320000 &   80000 &  800000&  1200000 &   1.95 &   2.26 &   0.14 &   0.07 &   0.79 &   6.25 \\
s47\_z15D   &  320000 &   80000 &  800000&  1200000 &   2.00 &   2.25 &   0.14 &   0.07 &   0.79 &   6.24 \\
s47\_z20D   &  320000 &   80000 &  800000&  1200000 &   2.05 &   2.27 &   0.14 &   0.07 &   0.79 &   6.25 \\
s57\_z05D   &  320000 &   80000 &  800000&  1200000 &   1.92 &   2.32 &   0.14 &   0.07 &   0.80 &   6.24 \\
s57\_z10D   &  320000 &   80000 &  800000&  1200000 &   1.95 &   2.26 &   0.14 &   0.07 &   0.79 &   6.25 \\
s57\_z15D   &  320000 &   80000 &  800000&  1200000 &   2.00 &   2.25 &   0.14 &   0.07 &   0.79 &   6.24 \\
s57\_z20D   &  320000 &   80000 &  800000&  1200000 &   2.05 &   2.27 &   0.14 &   0.07 &   0.79 &   6.25 \\
s27\_z05X   & 2133333 &  533333 & 5333334&  8000000 &   1.92 &   2.32 &   0.14 &   0.07 &   0.80 &   6.37 \\
s27\_z10X   & 2133333 &  533333 & 5333334&  8000000 &   1.95 &   2.26 &   0.14 &   0.07 &   0.79 &   6.35 \\
s27\_z15X   & 2133333 &  533333 & 5333334&  8000000 &   2.00 &   2.25 &   0.14 &   0.07 &   0.79 &   6.35 \\
s37\_z05X   & 2133333 &  533333 & 5333334&  8000000 &   1.92 &   2.32 &   0.14 &   0.07 &   0.80 &   6.37 \\
s37\_z10X   & 2133333 &  533333 & 5333334&  8000000 &   1.95 &   2.26 &   0.14 &   0.07 &   0.79 &   6.35 \\
s37\_z15X   & 2133333 &  533333 & 5333334&  8000000 &   2.00 &   2.25 &   0.14 &   0.07 &   0.79 &   6.35 \\
s37\_z10M   &  320000 &   80000 &  800000&  1200000 &   1.09 &   0.72 &   0.33 &   0.09 &   0.59 &   4.44 \\
s37\_z10MS  &  320000 &   87449 &  577113&   984562 &   1.00 &   1.00 &   0.33 &   0.09 &   0.59 &   4.44 \\
s37\_z10MX  & 2133333 &  533333 & 5333334&  8000000 &   1.09 &   0.72 &   0.33 &   0.09 &   0.58 &   4.40 \\
s37\_z10MXS & 2133333 &  586831 & 3775591&  6495755 &   1.00 &   1.01 &   0.32 &   0.09 &   0.59 &   4.48  \\\hline
\tabnotetext{ }{\scriptsize The first column is the name of the model which 
	indicates some initial conditions ($\sigma_{R,0}$ and $z_d$, see Figure \ref{modelstab02}).  
	The number of particles of each component are given in columns 2 to 4, 
	and column 5 shows the total number of particles for the system.  
	Column 6 gives the mass ratio between bulge and disk particles, while
	column 7 gives the mass ratio between halo and disk particles.
	Columns 8 to 10 give the total mass ratios of the components. 
	Finally, column 11 gives the total mass of the model $M_G$.}
\end{tabular}
%}
\label{models_tab03}
\end{center}
\end{table*}

\begin{figure}
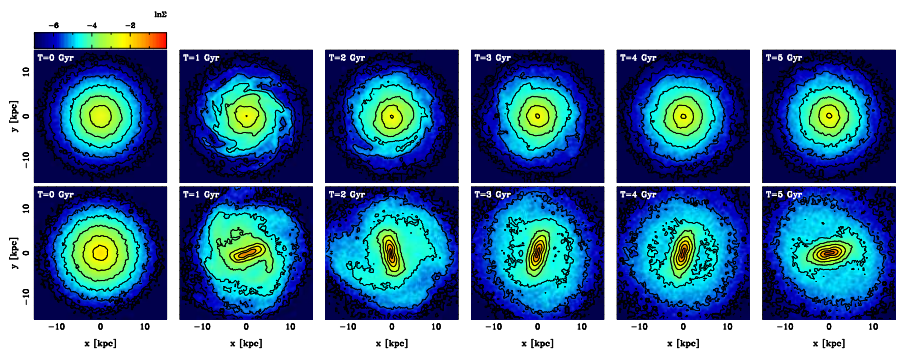

\centering
    \includegraphics[angle=-90,width=\columnwidth]{pgp_snap2_D12}\\
    \includegraphics[angle=-90,width=\columnwidth]{pgp_snap2_D12_B}
    \caption{Some snapshots of the models s27\_z10D (top row) and s37\_z10M (bottom row).
	    The model with a light disk forms spiral structures which fade out 
	    after some rotations. The model with a heavier disk forms a bar,
	    and this bar is maintained throughout the entire evolution. The time
	    in Gigayears is given in the bottom left corner.}
    \label{plotsxy}
\end{figure}

\subsection{Temporal evolution of the models}

The simulations performed in this work employ the N-body code
\mbox{gyrfalcON}, based on \citet{2000AGM....17..C01D,
2002JCoPh.179...27D} force solver \mbox{falcON}
(force algorithm with complexity) and the NEMO package 
\citep{1995ASPC...77..398T}. As tree-codes, \mbox{falcON} begins 
by building a tree of cells at each time-step, then determines
the potential of the system using multipole expansion for the cells
and finally, exploits the similarity of the force from a distance cell
upon cells that are close to each other. 

In all our simulations, we used a softening parameter 
$\varepsilon=0.05$ and an opening angle $\theta=0.5$. 
With these parameters, we ensure that the 
energy conservation is better than $10^{-4}$.  
The models were evolved from 0 to 5 Gigayears. 
Figure \ref{plotsxy} shows six snapshots 
for models s27\_z10D (top row) and s37\_z10M (bottom row). 
The s27\_z10D model forms transient spiral structures 
that fade out after some rotations, and the s37\_z10M 
model forms a bar structure that is maintained 
throughout the entire evolution.

\begin{figure}
\centering
 \includegraphics[angle=-90,width=\columnwidth]{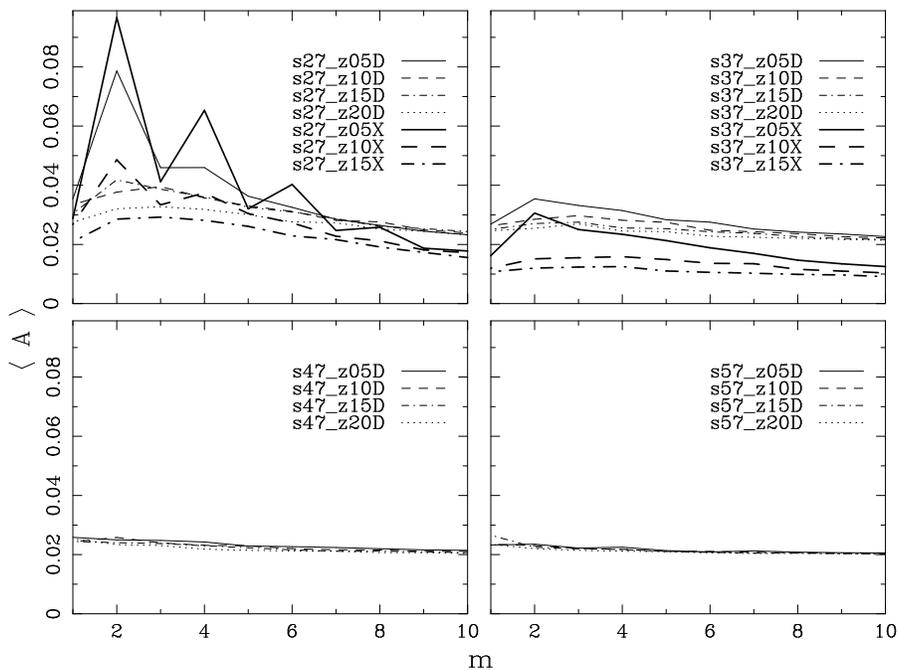}
 \caption{The average amplitude $<A>$ for modes $m$ 1 to 10 for all unbarred models.
	 The upper left panel shows the models with the lowest velocity 
         dispersion (cold models) and the bottom right panel shows models with the highest velocity 
         dispersion (hot models). While the coldest models generate strongest structures 
         at mode $m=2$, $m=3$ and $m=4$, which is evidence multi-armed structures,  
         the hottest models do not generate structures.}
 \label{suma_TF1D}
\end{figure}

\begin{figure}
\centering
 \includegraphics[angle=-90,width=\columnwidth]{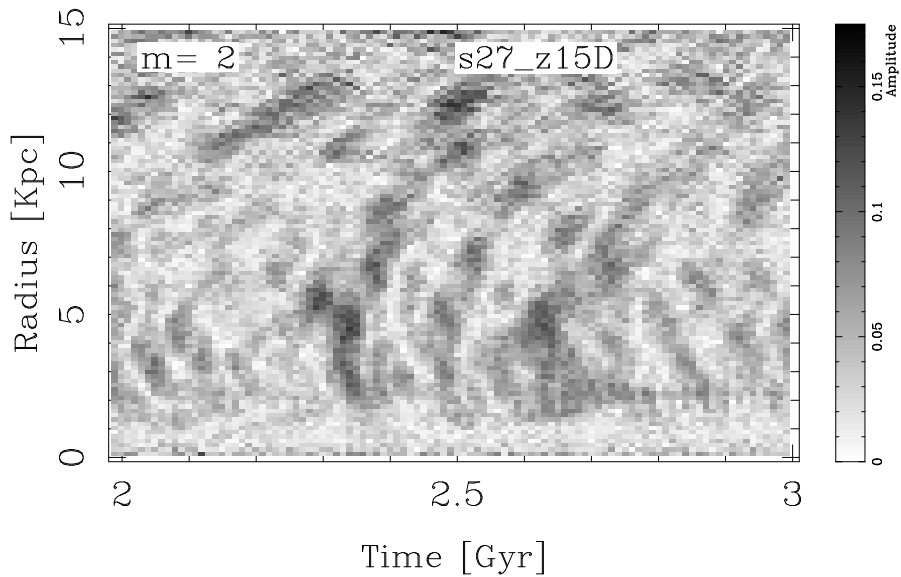}
 \caption{Zoom of the FT1D for model s27\_z15D for $m=2$ as a function of radius and time.
	 We can notice that the structures appear where the Toomre parameter $Q$ is minimum
         (see Figure \ref{curverotation}).}
 \label{zoom_ft1d}
\end{figure}

\subsection{Analysis of Models}
\label{sec_analysis}

We studied the time evolution of the models using one and two
dimensional Fourier Transform (FT1D and FT2D, respectively). 
High amplitude in the FT1D allows us to identify the region of
disk where the perturbation is strong and the FT2D allows
to get the pitch angle of the spiral structures.
Thus, performing FT1D and FT2D over each
snapshot allows showing how these modes (the
structures) and their pitch angles evolve during
the simulation. Moreover, the phase of the FT1D
coefficients gives information about the pattern
speed of the structures (bar and spirals). We
use the bar pattern speed to study the orbits of
the particles in the bar reference frame.

In order to implement the FT1D, we divided 
the disk into 100 rings for each given 
time $t$ (snapshot). Then, the FT1D is performed 
for each ring ($R=1$ to $100$) as follows:

\begin{equation}
A_{R}(m)=\frac{1}{D}\sum_{j=1}^{N_R} d_j e^{-im\theta_j}
\label{eqtf1d}
\end{equation}

\noindent where $D=\sum_{j=1}^{N_R} d_j$, $\theta_j$
is the azimuthal position of the $j$-th particle,
$N_R$ is the number of particles
for a given ring, $d_j$ is the weight of the $j$-th particle  
(in this case, we use the mass of each particle), and $m$ is the
azimuthal frequency.
We use this equation to get the Fourier coefficients for each 
mode $m$ and for each ring at each time step. 

The FT2D method is applied to the distribution 
of the disk particles as described in 
\citet{1992A&AS...93..469P} and references therein.
We applied the FT2D in an annulus with a minimum radius equal to 4.5 kpc
and a maximum radius of 15 kpc. The FT2D method is 
implemented by using a logarithmic 
spiral basis, $r=r_0\exp\left(-\frac{m}{p}\theta\right)$; 
where $m$ is a number of arms,
and $p$ is related to the pitch angle $P$ of the spiral structure by 
$\tan P =-m/p$. The discrete FT2D is given by the equation:

\begin{equation}
A(p,m)=\frac{1}{D}\sum_{j=1}^{N_A}d_j e^{-i(pu_j+m\theta_j)}
\label{eqtf2d}
\end{equation}

Here, $D=\sum_{j=1}^{N_A} d_j$, $N_A$ is number 
of particles in the annulus, $d_j$ is the weight 
of the $j$-th particle, $u_j=\ln r_j$, $r_j$ and 
$\theta_j$ are the polar coordinates of the
$j$-th particle. Thus, we applied the FT2D 
at each time step in order to obtain 
values of amplitudes A(p,m) for each $p$ and $m$.
These results allow us to analyze the growth of the spiral structures
and the evolution of their pitch angles.

\section{RESULTS AND DISCUSSION}
\label{results}

\begin{figure}
\centering
 \includegraphics[angle=-90,width=\columnwidth]{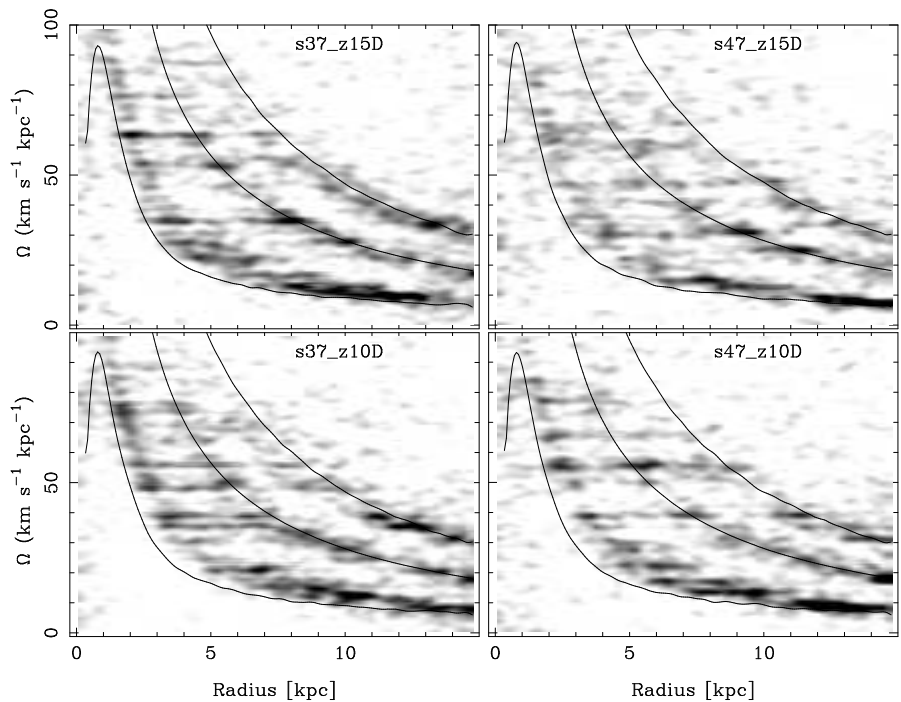}
 \caption{Spectrograms given the pattern speeds of 
 	      the structures for some unbarred models with 
          1.2M particles for the $m=2$ mode.  
          The curves are $\Omega+\kappa/2$, $\Omega$ and
          $\Omega-\kappa/2$.
          We note that the pattern speeds of the structures 
          (dark regions) are well confined between the inner and the
          outer Lindblad resonances.}
 \label{angular_velocities}
\end{figure}

\begin{figure}
\centering
 \includegraphics[angle=-90,width=\columnwidth]{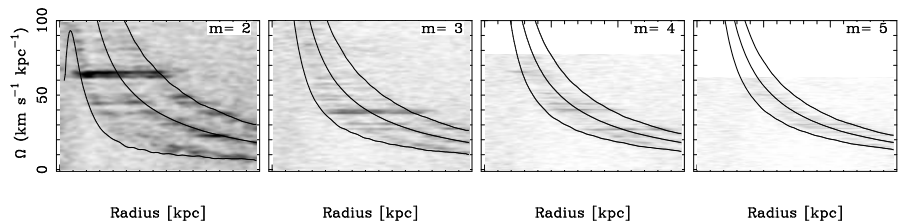}
 \caption{Pattern speeds for the model s27\_z15D for different $m$ modes.
          The curves are $\Omega$ and $\Omega\pm\kappa/m$ corresponding
	  to the main resonances. Dark regions correspond to the pattern speed of the 
          structures. It is clear that this model develops multi-armed spiral structure.
	  The pattern speeds of the structures for all $m$ are well located between
          their main resonances.}
 \label{pattern_speedm}
\end{figure}

\begin{figure}
\centering
    \includegraphics[width=\columnwidth]{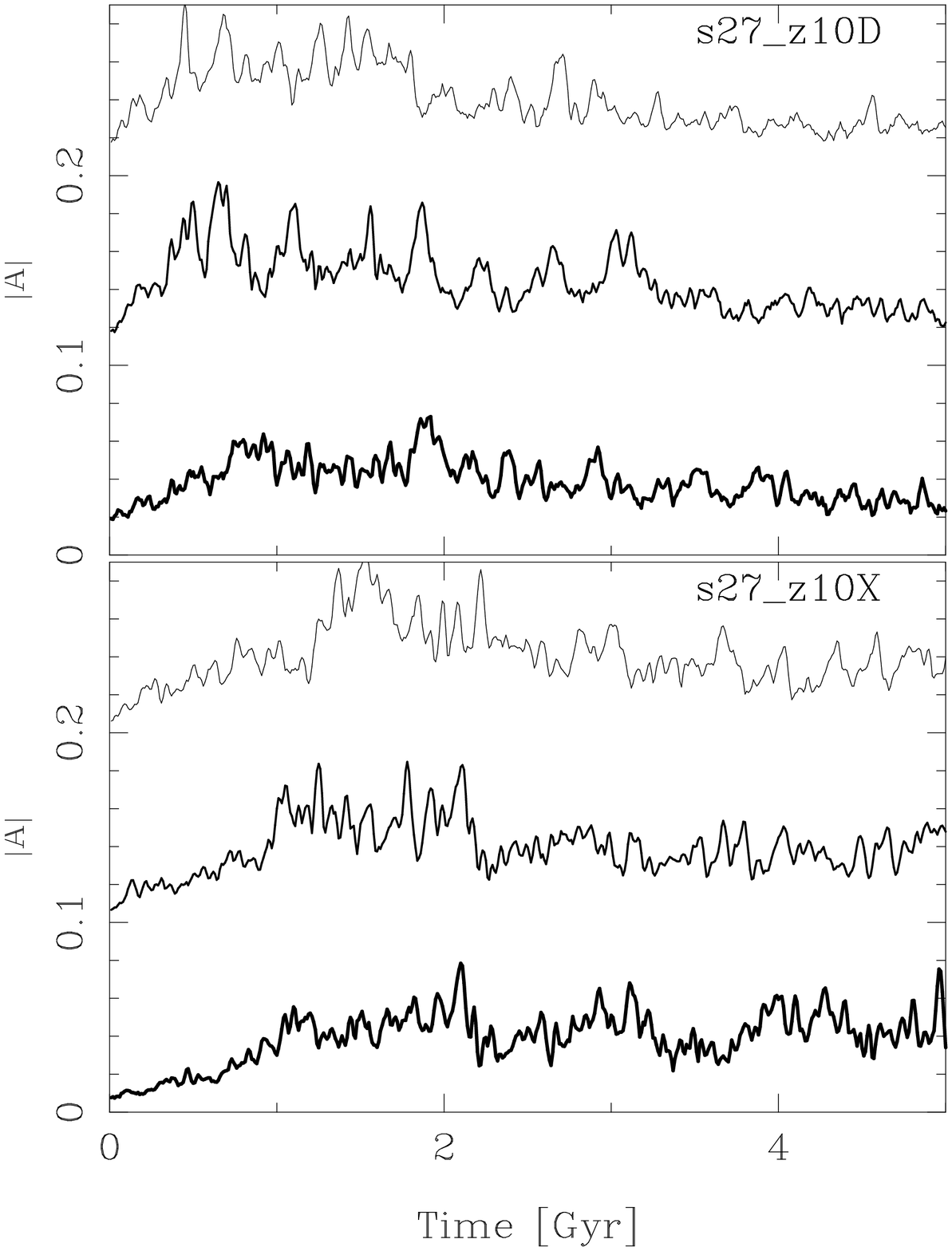}
    \caption{These plots show the average amplitude $|A|$ for mode $m=2$, $m=3$, and $m=4$.
	    We add 0.1 and 0.2 to the $m=3$ and $m=4$ for clarity.
	    The mean was calculated in the radial range 3.75 to 12.75 kpc.
             The thickest line is for mode $m=2$, the intermediate thick line is for mode
             $m=3$, and the thinnest line is for mode $m=4$. The upper panel shows the amplitudes for 
             the model with 1.2 million particles, and the bottom panel shows the amplitudes for
             the model with 8 million particles. We can see that these models generate multi-armed structures.
             Furthermore, we observe that the model with low mass resolution generates 
             structures earlier than the model with high mass resolution.}
    \label{ft1dm2_8million}
\end{figure}

\subsection{Unbarred models}
\label{mmodels}

In this section, we will present the calculations of the 
Fourier Transform methods (FT1D and FT2D); 
for the models with 1.2 and 8 million particles
which do not form a bar.

\subsubsection{FT1D for unbarred models}\label{s3}

We summarize the results of FT1D amplitude 
for all unbarred models in Figure \ref{suma_TF1D}. 
The mean amplitude of a given mode $m$ 
for each model was computed for all times and radii.
In this Figure, the upper left panel shows the cold models, 
while the bottom right panel shows the hot ones.
We have found that the cold models get the strongest structures 
at mode $m=2$, and there are also some strong 
structures at modes $m=3$ and $m=4$; 
in fact, models s27\_z05D and s27\_z05X 
have a predominant mode at $m=2$, 
due to the later formation of an inner oval and spirals. 
We conclude that the cold models are able to develop multi-armed
structures, i.e. spiral structures with different azimuthal 
frequency number ($m=2$, $m=3$, $m=4$), while
it is clear that the hot models are unable 
to develop strong spiral structures
with any azimuthal frequency.

In order to recognize where and when a $m$ mode is amplified,
we plot the Fourier coefficients $A_{R}(m)$ 
in gray-scale as a function of radius and time. 
The Figure \ref{zoom_ft1d} is an example of these plots.
This plot shows the amplitude of the FT1D for $m=2$
between 2 to 3 Gigayears for the s27\_z15D model.
The regions that have higher amplitudes (black stains) 
depict the growth of the structures. Especially in this graph, 
we can notice the evolution of a $m=2$ structure 
appearing around a radial distance of $5-6$ kpc. 
It worth to note that at those distances 
the Toomre parameter $Q$ is minimal. From there, 
the higher amplitudes move towards the inner 
and the outer parts of the disk.

\begin{figure}
 \includegraphics[angle=-90,width=\columnwidth]{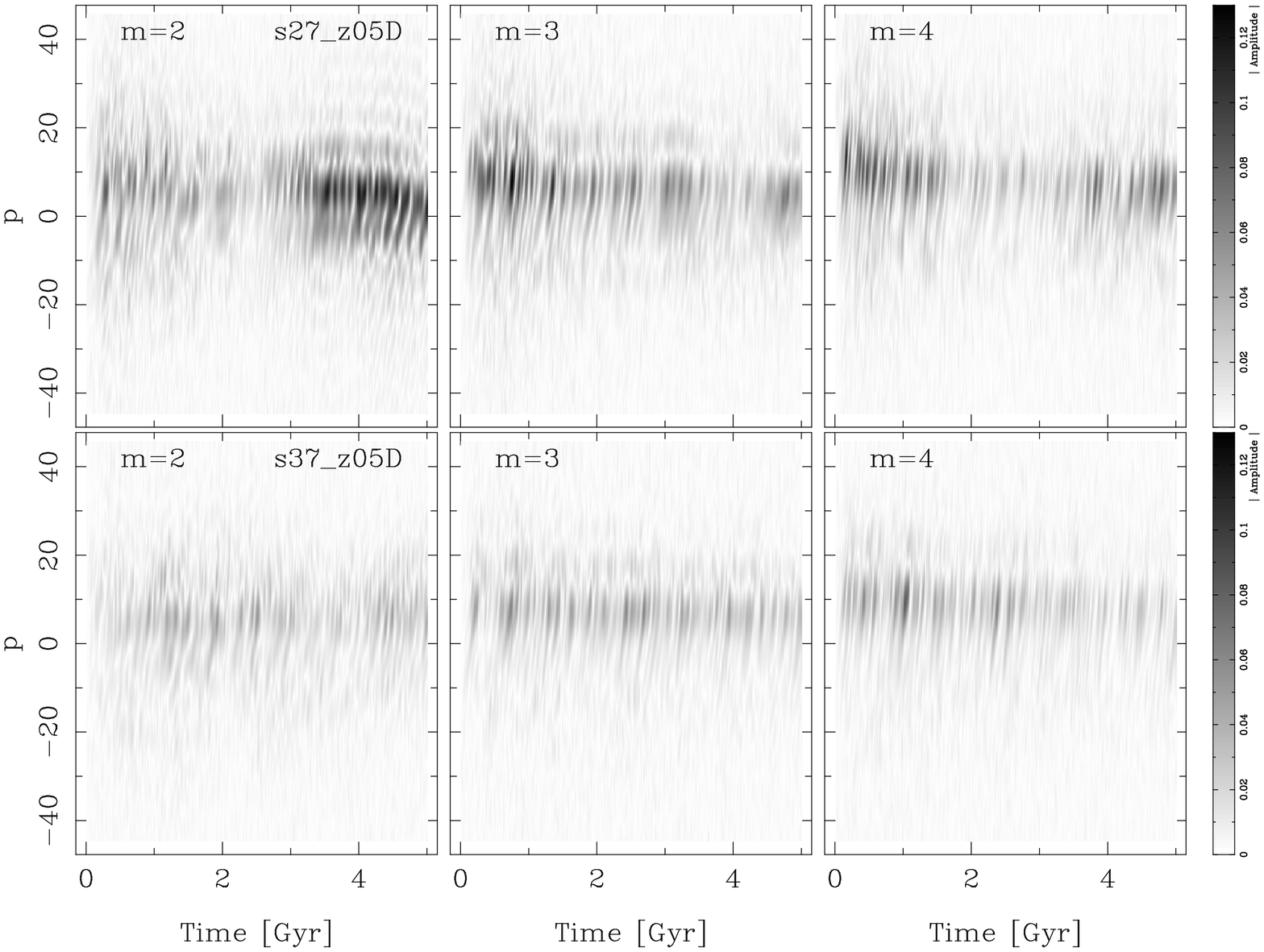}
 \caption{The amplitudes of the FT2D for $m=2$ for two of our unbarred models.
          Leading structures correspond to $p<0$, and trailing ones
          correspond to $p>0$. We can see how the amplitudes 
          oscillated with time; these oscillations are due to the superposition
          of waves with different pattern speeds.}
 \label{ft2dm2}
\end{figure}

All our unbarred models follow this behavior. 
Furthermore, we can notice that the perturbation begins 
to be amplified where the stability criterion 
is minimum around $5-6$ kpc (see Figure \ref{curverotation}).
It is known that models with low values of 
the Toomre parameter ($1<Q<1.2$), corresponding
to cold models \citep{1987gady.book.....B}, have the capability to 
develop non-axisymmetric structures via swing amplification 
while  models with high values of the 
Toomre parameter ($Q>2$), corresponding to hot 
models, are unable to form structures in the linear regime.
Indeed, our cold models (lower $Q$) are able 
to amplify strong non-axisymmetric structures
(Figure \ref{suma_TF1D}).

Our next step is to measure the location and the pattern speed of the
structures. Figure  \ref{angular_velocities} shows spectrograms of the
$m=2$ perturbations as a function of radius for the entire evolution
of these bisymmetric structures for four of our models.
The darker areas correspond to the angular velocity of the
strongest structures that are evolving in the disk.
We can note that these structures are 
constrained to be located between the ILR and the OLR resonances
implying that there are different structures with different 
pattern speed in the same radius.
Additionally, Figure \ref{pattern_speedm} shows the pattern speeds
of the s27\_z15D model for 
different modes $m$ modes illustrating that
structures with different $m's$ can co-exist in the galactic disk.
Like in \citet{2011MNRAS.410.1637S}, \citet{2014ApJ...785..137S}, 
\citet{2014ASPC..480..145V} and \citet{2014arXiv1408.3670M},
these Figures suggest that a spiral structure is
the result of the coupling of structures of a $m$ mode
and different pattern speeds, but also 
they are coupling in different parts of the
disk. The general morphology of our modeled galaxies
is due then to the superposition of several different modes.

\begin{figure}
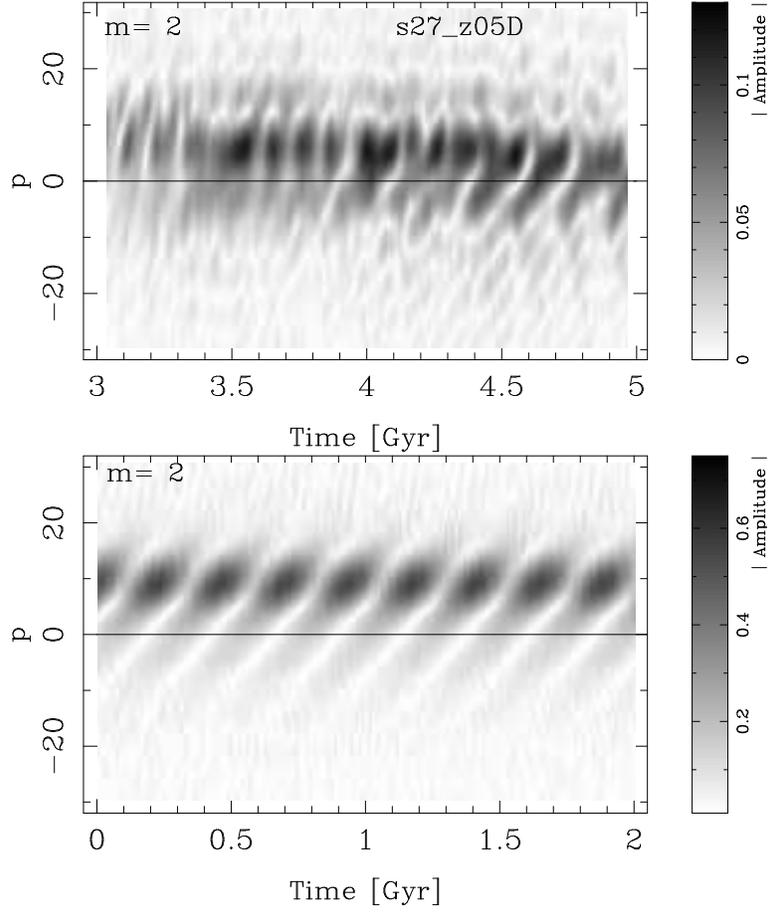

\centering
\includegraphics[scale=0.4,angle=-90]{TF2D_Am_m_p02_zumA}
\includegraphics[scale=0.4,angle=-90]{model_ln_m2_paper_TF2D}
 \caption{Upper panel: Zoom of FT2D at $m=2$ amplitudes for s27\_z05D
 	      model. Spiral structures with different pattern speeds 
 	      `beat' as waves to give these appearence of
          periodicity in the amplitudes. Bottom panel: The FT2D calculated using two spirals with
          different pitch angles and angular speeds. The superposition of these two spirals gives
          a Fourier Transform coefficientes behavior very similar to that one of ours modelled galaxies.}
 \label{zoom_ft2d}
\end{figure}

We also have evolved some models with larger
number of particles to investigate the growth
of the spirals. In this case, we calculated models
 with 8 million particles (see Figure \ref{modelstab02}),
i.e., we increased the number of particles by a factor of 6.6.
Figure \ref{ft1dm2_8million} shows the amplitude 
of the FT1D for different modes $m$ 
as a function of time. The upper panel shows 
the amplitude for the s27\_z10D model (model
with 1.2 million particles), and the bottom 
panel shows the amplitude for the s27\_z10X model
(same model with 8 million particles). 
These two panels are similar; this implies that 
the behavior of the model with more resolution 
is akin to the model with less resolution.

Therefore,  all unbarred models 
(with both higher and lower resolution) 
have similar behavior as we described before. 
However, the structures formed in these 
higher mass resolutions runs appeared a bit later and are weaker
than the structures formed in the less mass resolution runs.
It means that the heating of the disk for 
higher resolution models is weaker than 
the heating of the disk for lower resolution models.
The heating is due mainly to spiral activity and collisional
relaxation. The collisional relaxation is a minor effect compared
to the physically real collective heating caused by spirals.
Hence, the high-resolution models have less heating because
the spirals are weaker.

\begin{figure}[!t]
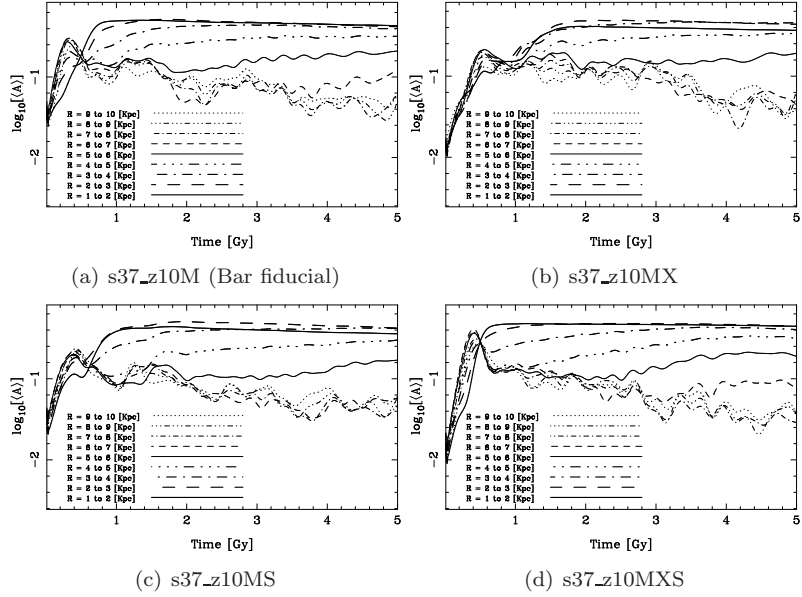

\centering
	\subfigure[s37\_z10M (Bar fiducial)]{\includegraphics[scale=0.3]{promedio_R123_FT1D}}
	\subfigure[s37\_z10MX]{\includegraphics[scale=0.3]{promedio_R123_s37_z10MX14_FT1D}}
	\subfigure[s37\_z10MS]{\includegraphics[scale=0.3]{promedio_R123_s37_z10MS14_FT1D}}
	\subfigure[s37\_z10MXS]{\includegraphics[scale=0.3]{promedio_R123_s37_z10M14XS_FT1D}}
	 \caption{The average of the FT1D amplitude for some radial ranges for the massive
	disk models for $m=2$. In these plots, one can note as the amplitude for inner radii
	grows and becomes more or less constant, representing the formation of the bar in the disk.
	The amplitude for outer radii oscillates representing the formation of transient 
	spiral structures.}
 \label{ft1d_media_barra}
\end{figure}

\citet{2011ApJ...730..109F}, \citet{2013ApJ...769L..24S} 
and \citet{2014ApJ...785..137S} explored models 
with a different number of particles $N$ in the disk.
\citet{2011ApJ...730..109F} found that the two-body
relaxation has effects on the heating of the disk only when the 
number of particles is quite small, while \citet{2014ApJ...785..137S} 
explained that the spiral activity causes 
the heating of the disk and the two-body 
relaxation is controlled by the softening
in the force computation.

We must note that our models with a high and
low number of particles have the same softening. 
The high-resolution models heat a bit later because
the two-body relaxation is less important than 
those with low resolution. The spiral activity
on high-resolution disk makes less heat them 
than low-resolution disk due to the smaller 
amplitudes of the perturbations.
We will strength this statement in section \ref{toomre}
where we discuss the evolution of the $Q$ parameter for all models.

\subsubsection{TF2D for unbarred models}\label{tf2d_normals}

We performed the FT2D on all our models, but for simplicity,
we only present in Figure \ref{ft2dm2} the results
for two of our models for modes $m=2$, $m=3$, and $m=4$. 
In this Figure, each panel depicts the Fourier coefficient 
(the amplitude) $|A(p,m)|$ on gray-scale 
as a function of time, and frequency $p$.
(see Section \ref{sec_analysis}). 
In these charts, the highest
amplitudes represents the spiral structures which are
appearing in the disk, and the black stains 
shows the evolution of these structures.

The oscillations we can see in the amplitudes of the spiral
patterns are a signature of superposition of
structures $-$ or modes $-$ with different values
of pitch angle and angular speeds. Such modes 
were already detected using the FT1D
spectrograms (see e.g., Figures 
\ref{angular_velocities} and \ref{pattern_speedm}).

In the upper panel of Figure \ref{zoom_ft2d}, 
we present a zoom of the FT2D
amplitude for mode $m=2$ between three 
and five Gigayears for the s27\_z05D model,
which is the coldest and thinnest model. 
In the bottom panel, we show a FT2D calculation
where we simulated 2 different spirals with
pitch angles 14 and 12.5 degrees. 
One spiral rotates with an angular speed
of 48 km/sec/kpc, while the other one 
rotates with $\Omega=35$ km/sec/kpc.
The superposition of these two spirals gives
a FT2D very similar to those we are measuring 
in our simulations. Then, once again
we emphasized that the Fourier analysis is showing that the
general morphology of our modeled galaxies is due to the
superposition of several different modes.

All our models, including models with high
and low mass resolution, follow this behavior
that was explained before. Moreover, we remark that
the coldest/thinnest models generate strong
spiral structures, while the hottest/thickest
models do not generate structures.
On the other hand, \citet{1984ApJ...282...61S} argued that the 
spiral arms in N-body simulations
generally fade out over time because 
the spiral arms heat the disc kinematically 
and cause the $Q$ to rise. Our models also show 
this behavior where the $Q$ parameter is increased. 
We show the increment of $Q$ in section \ref{toomre}.

\begin{figure}
\centering
\includegraphics[angle=-90,width=\columnwidth]{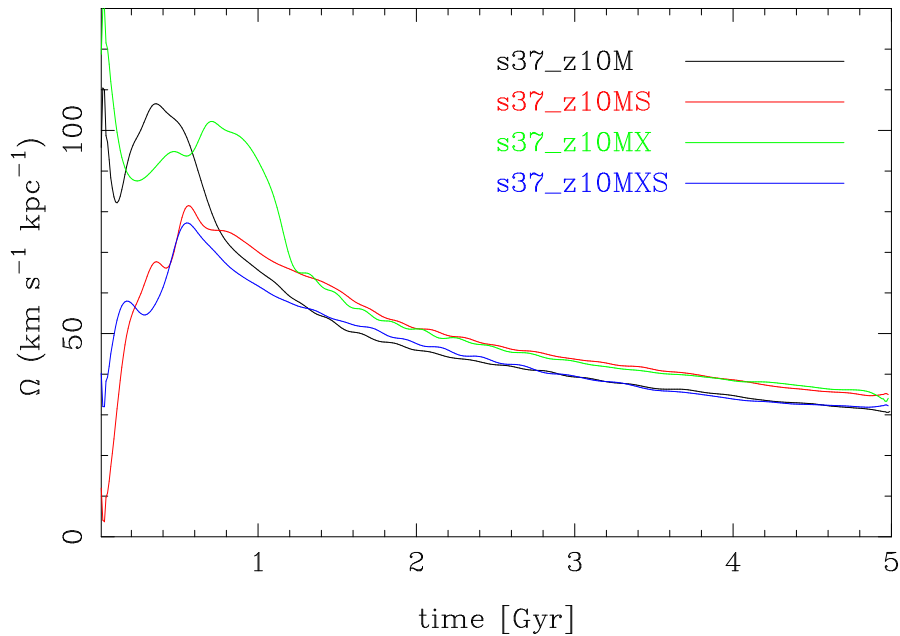}
  \caption{This Figure shows the pattern speed of the bar as a function of time 
      during the entire evolution for all barred models.
    The angular speed decreases in all models, from approximately $70-80$ km/sec/kpc at
    the first Gigayear to $30-35$ km/sec/kpc at end of our calculations, at 5 Gigayears.}
  \label{velo_angular}  
\end{figure}

\begin{figure}
\centering
	\includegraphics[width=\columnwidth]{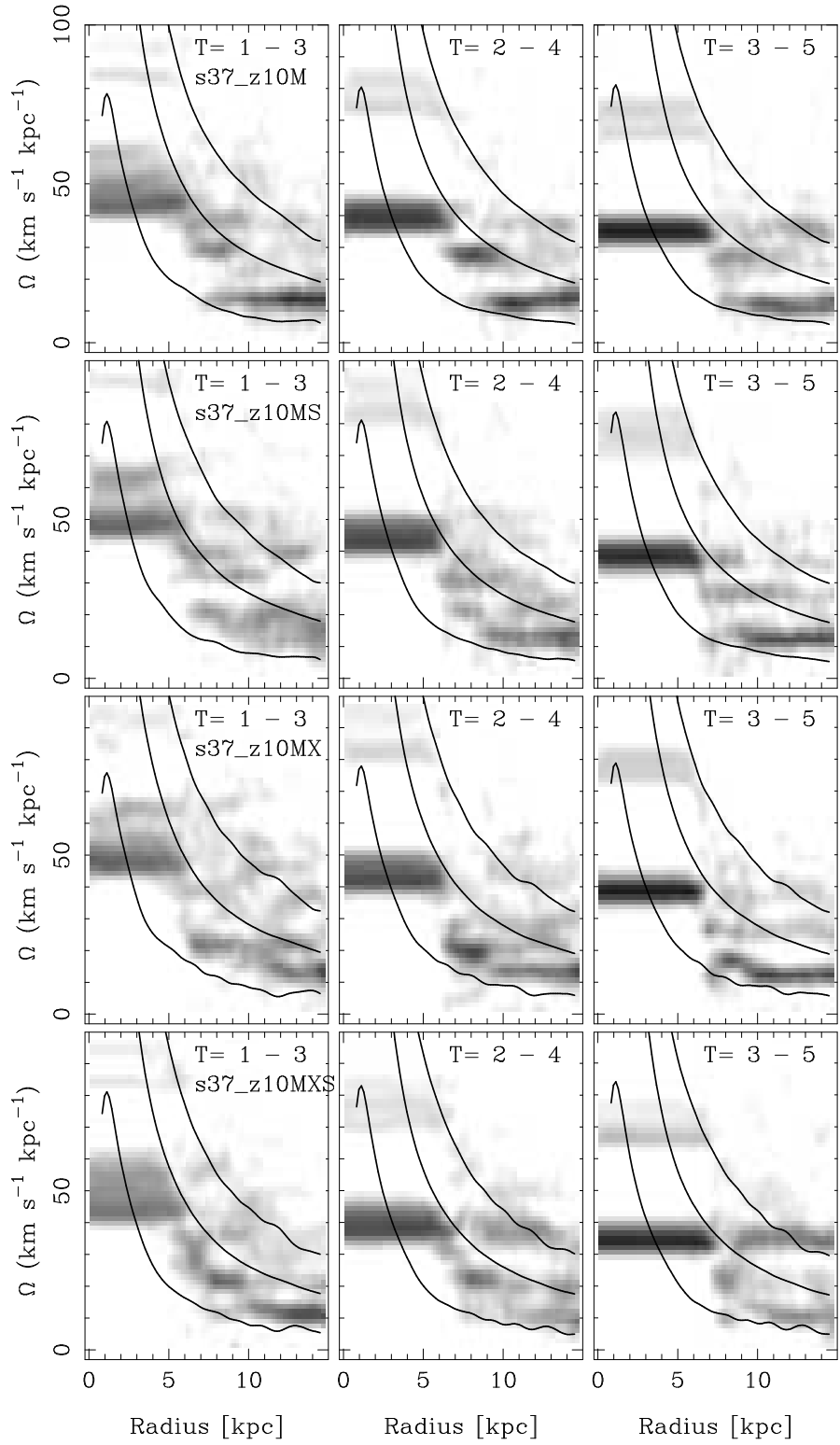}
	\caption{This Figure shows the spectrograms which were calculated for
        three-time intervals. We can see how the bar slowdown in its evolution. It is also
        clear that the spirals rotate slower than the bar (see text).}
 \label{three_spectograms}
\end{figure}

\subsection{Bar models}
\label{bmodels}
In this section, we will follow our analysis as we
performed in section \ref{mmodels}, but now on our barred models.
For these models, we also present an orbital analysis in order
to show the growth of the bar.

\subsubsection{TF1D for barred models}

In our sample of simulations, we ran models with a more
massive disk (see Figure \ref{modelstab02} 
and Table \ref{models_tab03}). These models clearly 
developed a bar feature because they had a quite low
initial $Q$ value at the beginning of their evolution 
due to the higher disk surface density.

We applied the FT1D in the same way as in 
the previous models to analyze the growth of the bar
and the spiral structures that are formed in the disk.
The results of the FT1D are shown in Figure
\ref{ft1d_media_barra}.
Each panel of this Figure shows the mean amplitude
$<A>$ (equation \ref{eqtf1d}) in logarithmic scale 
for $m=2$ as a function of time and radius 
for each model. These plots can be understood as
growing curves of the structures
that are being assembled. We can observe 
that the evolution of the amplitude is akin 
between all barred models, the main difference 
being the bar's growing rate.
For example, the amplitude in the inner disk
 between $1-2$ kpc for model s37\_z10M 
(called bar fiducial model) reach its maximum 
approximately at 0.9 Gyr of its evolution while
for model s37\_z10MX reach its maximum 
approximately at 1.4 Gyr. On the other hand, 
while the model s37\_z10MS reach its maximum 
approximately at 1 Gyr, the model s37\_z10MXS 
reach its maximum much earlier, at 0.6 Gyr.

In order to calculate the instantaneous bar pattern speed as a 
function of time, we calculate the angle $\theta$ of the bar
from the phase $\phi$ of FT1D for $m=2$  taking the mean value
of $\theta$ in a radial range from 0.15 to 2.25 Kpc, 
getting then a $\theta(t)$ curve. We then fit a 
straight line each 5 points on the $\theta(t)$ curve
and take its slope for the middle point 
as a result of pattern speed $\Omega_p(t)$. 
Figure \ref{velo_angular} shows $\Omega_p(t)$ 
for all barred models. It depicts the evolution of 
their instantaneous bar pattern speed evidencing 
that the bars appear with high angular velocity, 
and they slow down throughout the time.
\citet{2014MNRAS.438L..81A} showed that this 
decrease of the pattern speed could be 
explained as due to the angular momentum 
exchange within the galaxy or by the dynamical 
friction exerted by the halo on the bar.
In a forthcoming article 
(Valencia-Enr\'\i quez et al., in preparation)
we will present results focused in a large set of
isolated/interacting barred galaxy simulations 
to get insight on the processes which speed up or slow
down the bar.

In Figure  \ref{three_spectograms}, we present the $m=2$
spectrograms of these simulations, now calculating them for
three different time intervals ($1-3$, $2-4$, $3-5$ Gyr). 
This calculation allows us to compare the
bar pattern speed and the spirals pattern speed.
We can clearly observe the bar pattern speed in the inner part 
of the disk with high amplitude (dark horizontal line), 
and the pattern speed of the spirals in the outskirts of 
the disk with a bit less amplitude.

The spectrograms of all barred models in 
the Figure \ref{three_spectograms} show that 
the bar slows down more than the spiral. Furthermore,
the bar rotates around twice as fast as the spiral structures.
The spiral structure, as in the previous cases,
results from the superposition of several 
$m=2$ modes with different pattern speeds at different 
radial regions (see \citealt{apj_722_1_112} 
for an extensive discussion on the spiral$-$bar 
resonance overlap). The difference between barred 
and unbarred models is that in barred
models, there are more $m=2$ modes being
amplified, probably due to the influence of the strong fast
rotating central bar.

\begin{figure}
 \includegraphics[angle=-90,width=\columnwidth]{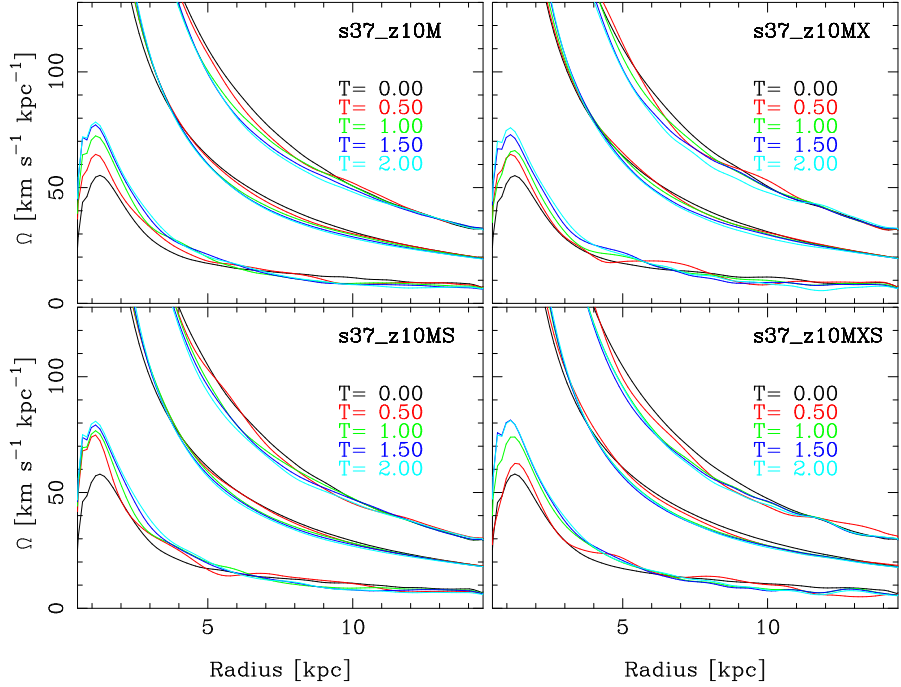}
  \caption{This Figure shows angular frequency $\Omega$ 
    (middle lines), $\Omega-\kappa/2$ (bottom lines), and $\Omega+\kappa/2$
    (upper lines) at different times for all bar models.}
  \label{resonancess}  
\end{figure}

Also, we show the main resonances curves 
to identify whether the bar has an ILR. 
These results are shown in Figure \ref{resonancess}.
To get these curves, we obtain $\Omega$ from 
the circular velocity of the model, and the 
main resonances from the equations  $\Omega \pm \kappa/2$
where $\kappa$ is the epicycle frequency
($\kappa^2=R\frac{d\Omega^2}{dR}+4\Omega^2$).
This Figure shows the $\Omega$ and the main resonance 
for all barred models from 0 to 2 Gigayears. 
We can observe that all our models have a 
high ILR  in the inner region of the disk because of 
the presence of a strong bulge, and it is rising 
due to the increasing of the density in the center 
of the disk across their evolution. 
Yet, we must note that the bar appears with very 
high angular velocity (see Figure \ref{velo_angular}), 
higher than the top of the ILR at the beginning
of the simulation (see the ILR in Figure \ref{resonancess} 
at $T=0$ drawing with black line) implying 
the bar is probably formed in the linear regime
\citep{1986MNRAS.221..213A}, but after some time 
the bar slows down quickly and the top of the ILR 
overtakes the angular velocity of the bar. 
It happens probably due to the exchange of angular momentum
(\citealt{2003LNP...626..313A} and references therein). The bar
finally evolves having an ILR \citep{2015PASJ...67...63O}.

\subsubsection{TF2D for barred models}

\begin{figure*}
 \includegraphics[angle=-90,width=\columnwidth]{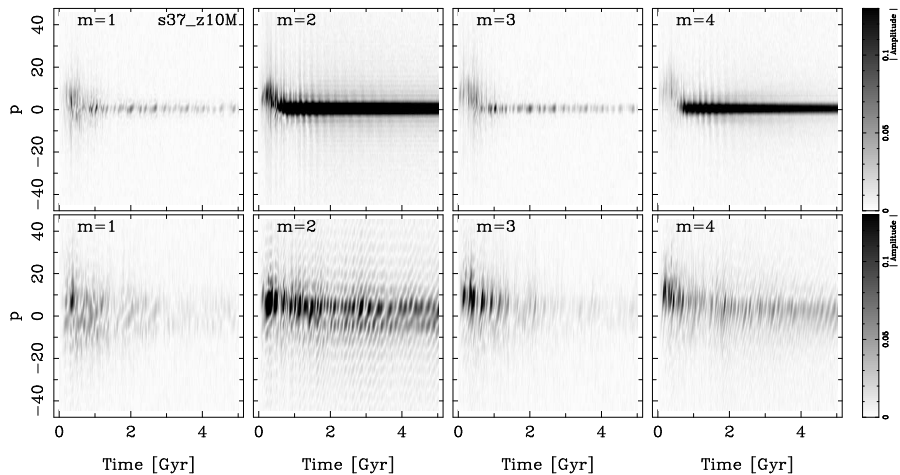}
 \caption{Upper panels show the amplitude of the FT2D from $m=1$ to $m=4$ for the fiducial 
	  bar model. The FT2D was calculated at an inner radial region $0.1<R<4.5$ kpc.
          The maximum amplitude being found for the Fourier coefficient $A(p=0, m=2)$
          is a clear sign of the presence of a bar.
          Bottom panels show the amplitude of the FT2D for the same modes $m$, 
          but for the outer radial region $4.5<R<15$ kpc.
	  In the barred cases, the
	  spiral structures are stronger than in the unbarred ones. For the
	  outer radial region, higher azimuthal frequencies are triggered
	  at the time of the bar formation.}
 \label{ft2d_barra1}
\end{figure*}

For barred models which have a heavier disk 
(green box in Figure \ref{modelstab02}),
the FT2D was separately computed for both 
bar ($0.1<R<4.5$ kpc), and disk ($4.5<R<15$ kpc) regions. 
Figure \ref{ft2d_barra1} shows the Fourier 
coefficients of FT2D for the fiducial model; 
the results of the FT2D for the other barred models 
are very similar. The upper panels of this Figure clearly
show the presence of a bar structure; 
there is very high amplitude for $m=2$ and $p=0$. 
The bottom panels show the spiral arms 
that are growing in the disk.

\begin{figure}
  \includegraphics[angle=-90,width=\columnwidth]{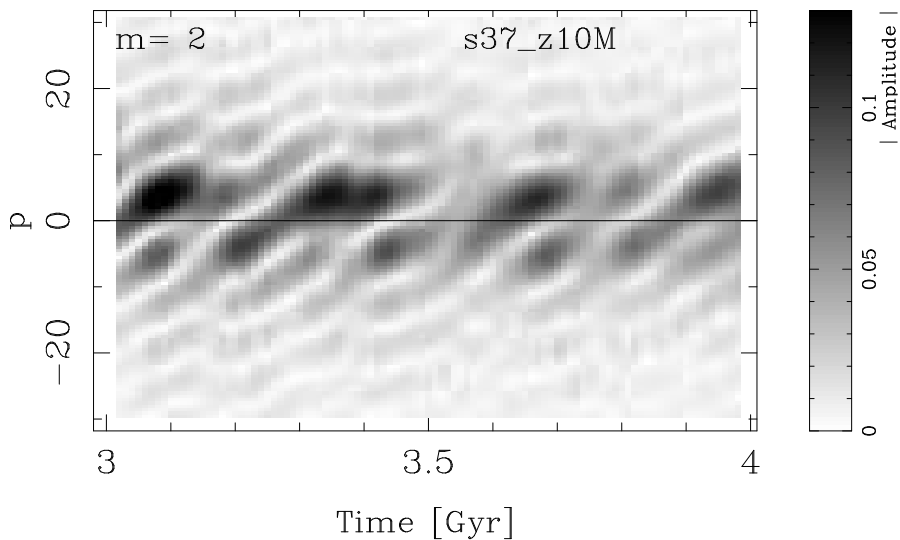}
  \caption{Zoom of the FT2D amplitude at $m=2$ mode for the barred model calculated 
           in the radial interval $4.5<R<15$ kpc. We can observe 
           that both leading ($p<0$, with less amplitude) and 
           trailing spiral structures ($p>0$, with higher amplitude) appear
           for a given time due to the superposition of waves with different
	   pitch angles and angular speeds.
%	   These spiral structures 
%           present a cyclical behavior (approximately 100 million years) like unbarred models. 
           Besides, we can note the amplitude of these spirals is higher compared to those of the unbarred models
           (Figure \ref{zoom_ft2d}).}
 \label{ft2dm2_zoom_bar}
\end{figure}

Similarly to section \ref{tf2d_normals}, 
we zoom of the $A(p,m=2)$ coefficients
in the interval between 3 and 4 Gigayears 
for the disk ($4.5<R<15$ kpc) region.
These results are shown in Figure \ref{ft2dm2_zoom_bar}. 
In this Figure, we note that the spiral structures 
get stronger due to the presence of a bar 
\citep{butaetal2009} compared to the lighter 
disk simulation (see  Figure \ref{ft2dm2}),
All in all, the spiral waves in this barred 
model have a similar behavior to the unbarred models.
Which is a clear evidence of effects of the 
coupling of the spiral structures, now
reinforced with the central bar.

%\begin{figure*}
 %\includegraphics[scale=0.9]{FT2D_mkd_s37_z10M14_s37_z10M14.pdf}
 %\caption{.}
 %\label{ft2d_barra2}
%\end{figure*}

\subsubsection{Orbits for the barred models}

\begin{figure*}
  \includegraphics[width=\columnwidth]{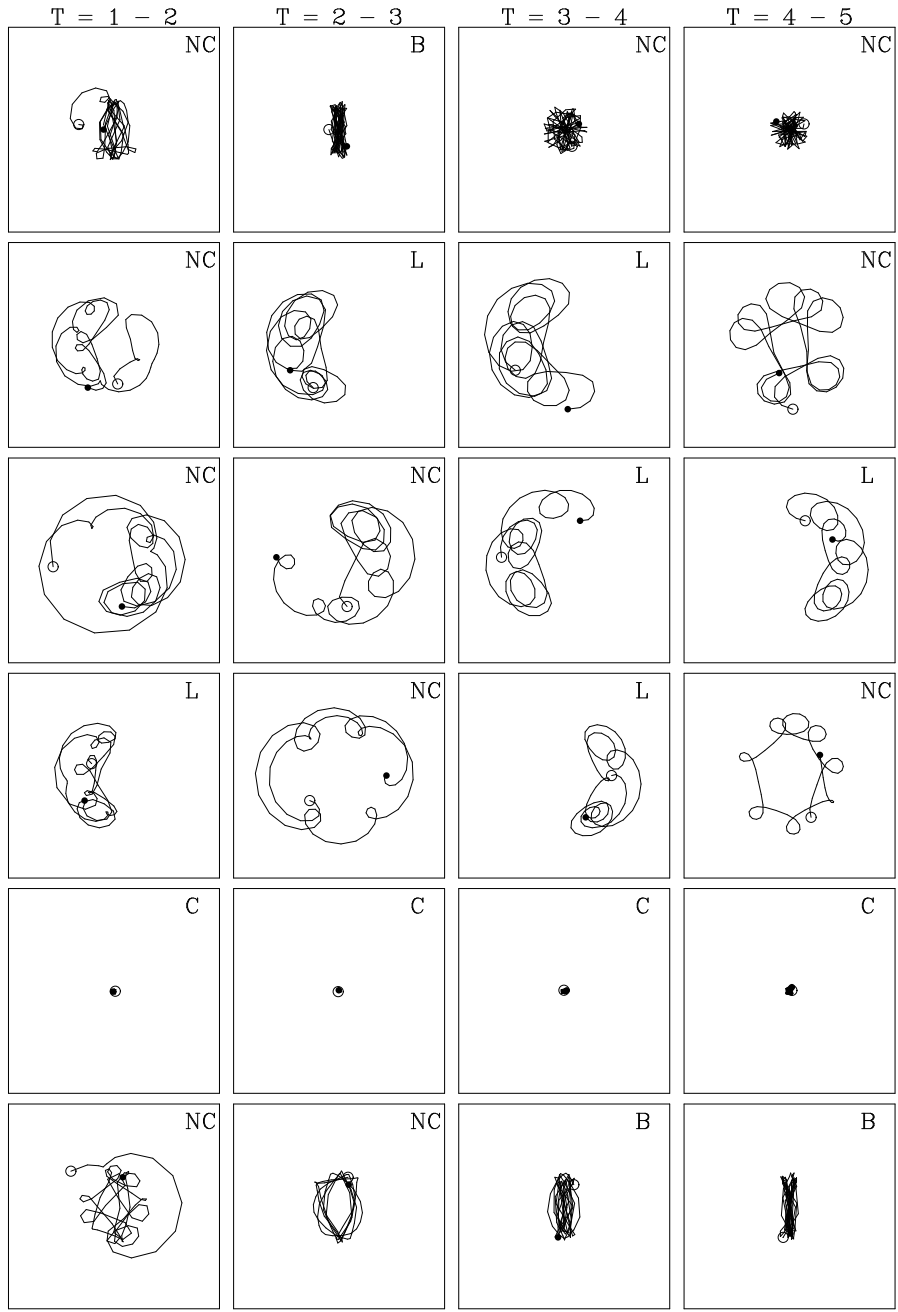}\\
  \caption{Morphological evolution of six particles of the s37\_z10M model.
      At each row, we plot the orbit of the same particle in 4 different time intervals.
      The orbit classification is given at the upper right corner. We can observe that a
      particle is confined to an orbit type morphology, but they can change
      during its entire evolution.}
 \label{classify}
\end{figure*}

We now focused our interest on an analysis of the 
disk stellar orbits in the bar reference frame.
For example, \citet{2011MNRAS.416..479C} 
proposed a complete analysis of the 
stellar orbits in test particles of 
gravitational potentials which are based
on tracing patterns in sequences of characters, 
which indicate sign changes of the Cartesian coordinates.
On the contrary, we propose a very simple 
geometrical method in which we analyze 
three morphological types of orbits 
in the bar reference frame. For illustration,  
we show in Figure \ref{classify} the morphological
evolution of six orbits for the fiducial barred model. 
We can clearly notice that a given particle is not
confined to one only type of morphology; they change 
their morphologies during their entire evolution
\citep{2012MNRAS.426L..46A}.

In order to make the analysis of the stellar orbits which
evolved with the bar formation, we graphically classify the
orbits of the disk particles in three primary morphological
types as follow: orbits that are concentrated
at the galactic center (compacts, C), orbits that are along
the bar (bar, B), and  orbits that are around 
the Lagrangian points $L_4$ and $L_5$ (loop orbits, L).
We developed a code to choose these type of orbits where 
we take our bar reference frame in a vertical position,
and we used the following criteria.  The criterion to
classify barred type orbits is $|y_{max}|>1.9|x_{max}|$ 
where $|y_{max}|$ is the maximum position value 
of a given particle in the vertical component 
(along of the bar) and $|x_{max}|$ is the maximum position
value of the same particle in the horizontal component
(perpendicular to the bar). The criterion to 
classify compact type orbits is $r_{max} \leq 0.5 \ kpc$, 
where $r$ is the maximum radius of a given particle. 
Finally, the criterion to classify loop type orbits
is that the particle orbit is confined 
to one side of the bar. Particles not following 
either of our three criteria are then unclassified (NC).
These criteria were used for all disk particles 
for each one Gigayear interval.

\begin{figure}
  \includegraphics[angle=-90,width=\columnwidth]{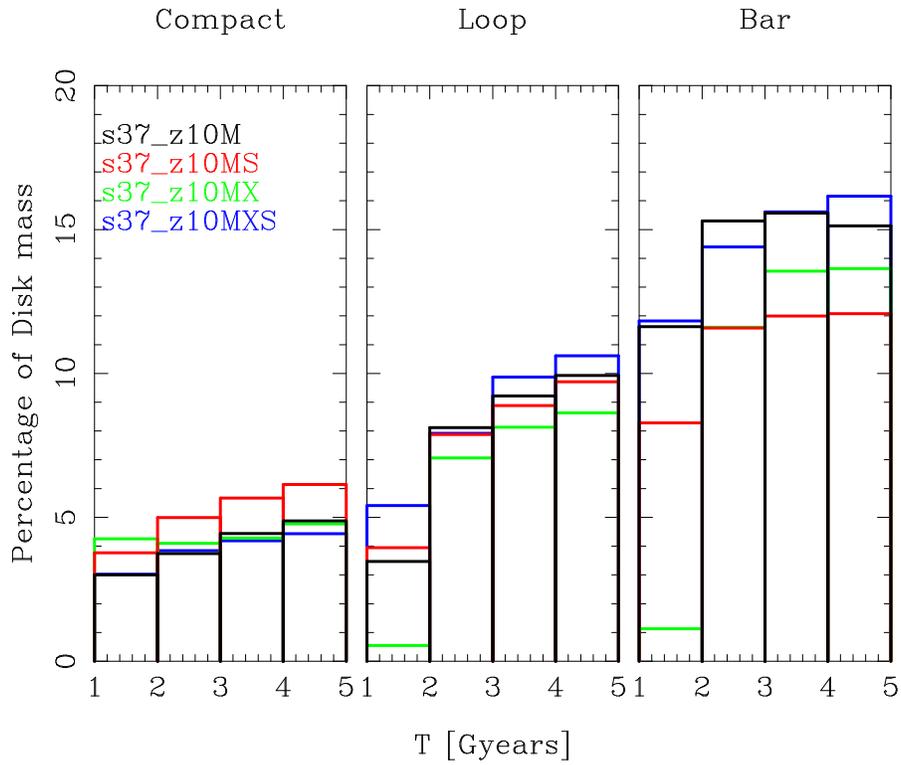}
  \caption{Morphological classification for all disk particles orbits at 4 times intervals.
           The left panel shows the compact type orbits, the middle panel 
           shows loop type orbits, and the right panel shows bar type orbits.
	   At the end of our calculations (5 Gigayears) these three type of orbits enclosed approximately
       one-third of the disk mass (see text).}
  \label{histogram_orbits}
\end{figure}

Furthermore, we have to keep in mind that the bar appears
around 1.5 Gigayears for the s37\_z10MX model, 
and around 1 Gigayear for other models.
Hence we made this orbital analysis 
from T=1 to T=5 Gigayear for all barred models.
These results are shown in Figure \ref{histogram_orbits}.
This Figure shows the statistical classification of
these three types of orbits which are weighted on
the percentage of disk mass. We can see in this Figure 
that the general behavior is akin to the all barred
models. For example, with the bar formation, 
the number of their compact orbits have a little growth,
and their loop orbits have a more pronounced growth. 
The number of their bar orbits remains more or less 
constant after the second Gigayear.

These models have the same initial conditions; the only
difference is the number of particles, which defines 
the mass of the particles. This difference
makes that the noise affects a little bit different
the evolution of the models \citep{2003ApJ...587..638S}. 
For example, we can find that the model s37\_z10MS 
has more compact orbits than the other models, 
the number of loop orbits is almost equal to the fiducial
models, and this forms approximately 4\% fewer bar orbits
than the fiducial model. The largest model (s37\_z10MX) 
keeps its compact orbits around 4\% across its entire
evolution, this model has 1.5\% fewer loop orbits 
and 3\% fewer bar orbits than the fiducial model 
afterward the second Gigayear.
Finally, we can see that the another one large model
s37\_z10MXS is the most similar to the fiducial model
because its differences in all type of orbits is around 1\%.

In general, the compact type orbits increase their number
up to  5\% of the disk mass, the loop type orbits 
increase their number to achieved around
10\% of the disk mass for all of the models, and 
the bar type orbits increase their number to attain around of
15\%, 13\%, 12\% and 16\%. for fiducial,  s37\_z10MX,
s37\_z10MS, and s37\_z10MXS models, respectively. 
Finally, we have found that these three geometrically
classified type of orbits, which are the ones
trapped in the Lagrangian points $L_3$, $L_4$, 
and $L_5$, encompass approximately one-third of 
the disk total mass for these bar models.

\begin{figure}
 \includegraphics[width=\columnwidth]{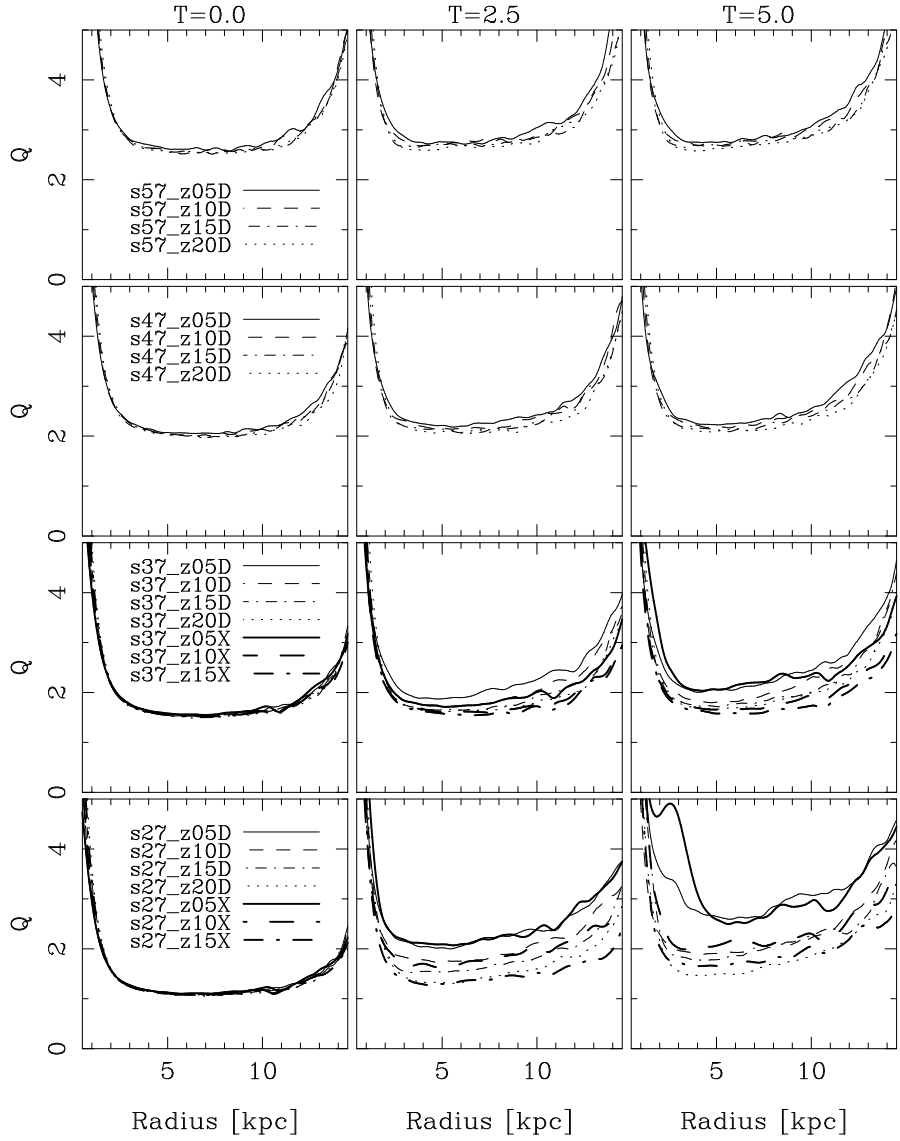}
 \caption{The evolution of the $Q$ parameter as a function of 
          radius for all unbarred models. The $Q$ parameter
  is shown at T=0, 2.5, and 5 Gigayears (see text).}
 \label{qparameter}
\end{figure}

\subsection{Toomre stability parameter $Q$}\label{toomre}

\begin{figure}
 \includegraphics[width=\columnwidth]{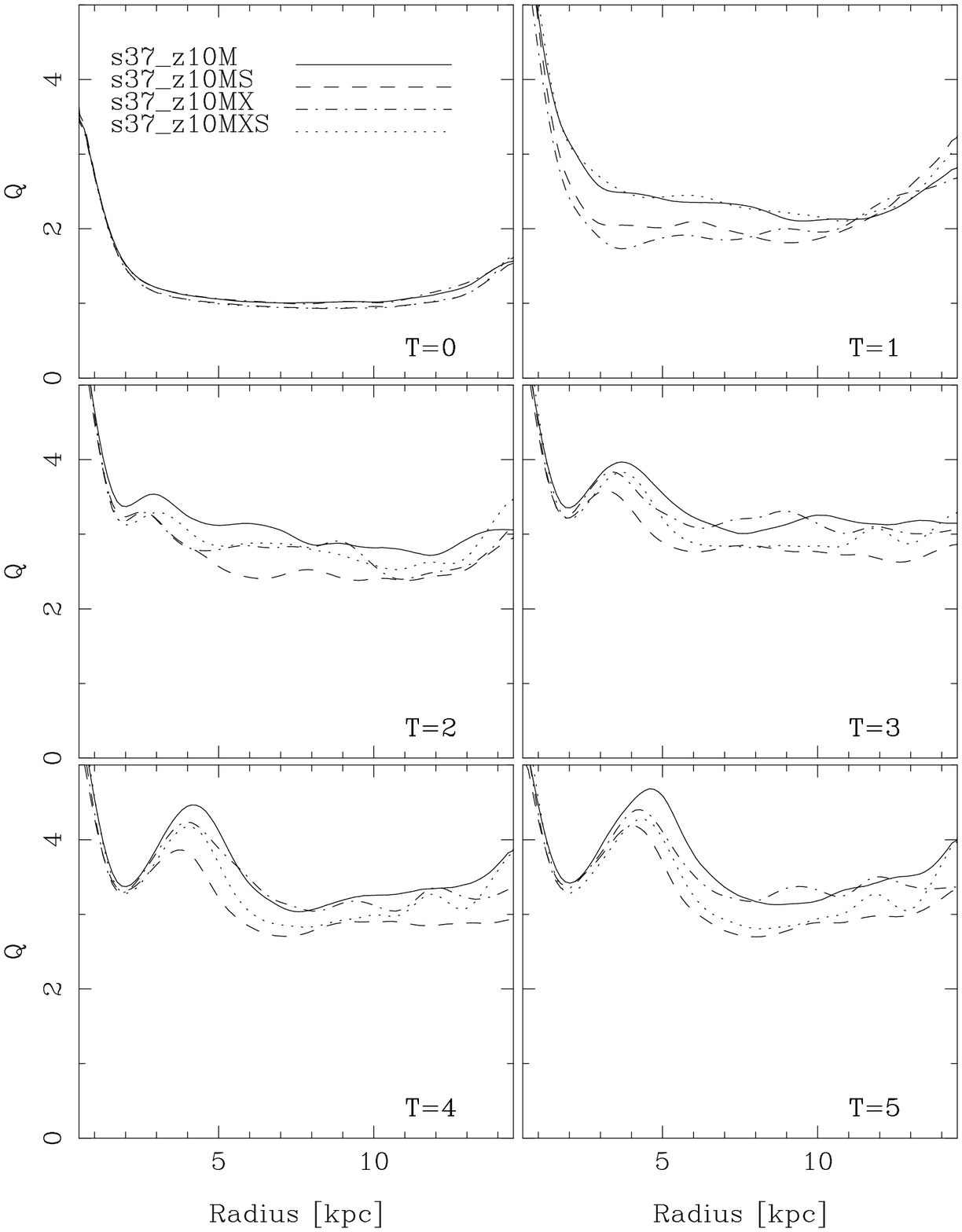}
 \caption{The Figure shows the evolution of the $Q$ parameter as 
   a function of radius for the barred models. The $Q$ parameter
    for all barred models have low values at the beginning of the simulation,
    but it increases considerably due to the bar and spirals formation.}
 \label{qparameter_bar}
\end{figure}

We already have discussed that the thickness 
of the disk $z_d$ affects the formation 
of non-axisymmetric structures. For example, Figure 
\ref{suma_TF1D} shows that the models with a very 
thin disk (e.g. s27\_z05D or s27\_z05X)
form stronger structures than models
with thicker disks demonstrating 
that the local instabilities of the disk also
depends on its thickness. 

It is already known that the stability of the disk
is measured by the Toomre parameter $Q$ in the standard first
order perturbation where the disk is considered 
like a very thin, self-gravitating disk
\citep{1987gady.book.....B,1996ssgd.book.....B}. 
However, as we said before, the thickness of the disk 
is also related with the stability of the disk, and so
it is convenient to study the stability of 
the disk ($Q$) for different values of the 
vertical scale height $z_d$.

The measure of the $Q$ as a function of radius at
three different times (0, 2.5, and 5 Gigayears) 
for unbarred models is shown in Figure 
\ref{qparameter} evidencing that the models with
low/high velocity dispersion have low/high
values of the parameter $Q$. Additionality,
we note that the hotter models, in upper panels, 
keep the parameter $Q$ approximately constant through 
the time while the initially colder models, 
in bottom panels, present an increment of $Q$. 
Furthermore, it depends strongly on the initial 
disk thickness $z_d$. For example, the increment 
of $Q$ for model s27\_z05D (or s27\_z05X) is more
conspicuous than the increment of $Q$ 
for model s27\_z20D which has a thicker disk. 
Hence, the stability of the disk should depend on 
the velocity dispersion in the $z$ component ($\sigma_z$),
the vertical scale height $z_d$, and mass of the disk. 
These results have also been discussed in
\citet{2009MNRAS.398.1027K}. 

We also measure the evolution of $Q$ parameter for barred
models; these results are shown in Figure
\ref{qparameter_bar}. We can see in this
Figure that the parameter $Q$ is lower at 
the beginning of the simulation, but
it increases quickly due to the bar and spiral formation.
However, as the bar reaches its saturation and evolves far from
the linear regime, the increment of $Q$ does not affect 
the bar strongly evolution and neither the maintenance
of the spiral structures.

Additionally, we have to claim that the models with 
few and high number of particles and same initial 
conditions heat the disk in similar way (see Figures
\ref{qparameter}, and \ref{qparameter_bar}).
\citet{2014ApJ...785..137S} explain that this 
behavior is due more to the spiral activity than 
to the two-body relaxation (\citealt{2011ApJ...730..109F}).
Moreover, it also depends on the softening parameter, if this
is quite small, then the two body-relaxation could be
important in the evolution of the model
\citep{2013ApJ...769L..24S}, and the disk is quickly heating.
See also \citep{1998AA...335..922R}
for an extensive study on the softening choice. 

\section{Summary and Conclusions}
\label{summary}

We performed a series of 3D fully self-consistent
N-body simulations with 1.2 and 8 million particles. 
The initial conditions were chosen to follow 
Kuijken-Dubinski models. In this work, 
we run a grid of models with different 
disk radial velocity dispersion $\sigma_R$, 
disk scale height $z_d$, number of particles $N$, 
and disk mass $M_D$. We analyzed the growth of spiral
structures by using one and two dimensional 
Fourier Transform (FT1D and FT2D). The FT's give 
the amplitude, the number of arms, and 
the pitch angle of a particular spiral structure.

The FT1D was a powerful tool in order to 
understand the growth of the spiral
structures. The results of the FT1D show us 
that the spiral structures emerge in 
the intermediate portion of the disk, 
where initially the $Q$ parameter reaches
its minimal value and these structures grow 
towards the outer parts of the disk with 
more intensity than they grow inwards, 
due to the steep increment of $Q$
towards smaller radii.

The plots of the FT2D amplitude as a function of
time and pitch angle show that {\it the general morphology
of our modeled galaxies is due to
the superposition of structures with 
different values of $p$, $m$, and angular velocity}.

We measured the angular velocity of all amplified patterns, and
the results show that these patterns 
are very well confined between the
main resonances given by the $\Omega \pm \kappa/m$ curves.
Moreover, we found that different structures with 
either different or same mode $m$ and frequency $p$,  
and different pattern speeds can evolve in the same region
of the disk at the same time.
Therefore, it is important to note that very
often two or three different spiral structures can
coexist in the same region of the disk.

\citet{1997A&A...322..442M} show signature 
of non-linear coupling between the bar
and spiral waves, or between spiral waves 
from modes $m=0$ to $m=4$. Another similar 
frame is present by \citet{2011MNRAS.410.1637S} 
where he showed that the bisymmetric spirals are not
a single long-lived pattern, but the superposition
of three or more waves can grow and decay with time. 
On the other hand, we have shown that not only
the spirals can overlap with different mode $m$, 
but also they  can overlap with 
 different frequency $p$. Thus, 
the general morphology of our modeled
galaxies is then due to the superposition of
structures with different values for $p$, and $m$,
i.e., different pitch angle and number of arms.

We have again to remark that the mass ratios 
$M_D/M_G$, $M_B/M_G$, $M_H/M_G$,
and the initial condition between models with few and high
number of particles are equals, but the mass of the particles
are different. Therefore, the evolution of these
models are affected by the noise 
(\citealt{2003ApJ...587..638S}, \citealt{2007MNRAS.375..425W}a,b).
For example, the bar in the model s37\_z10M is formed
earlier than in the models s37\_z10MX, probably due to the
same softening we used for all models. 
The noise is more incremented in models with 
few number of particles than those 
high number of particles. Therefore, it is clear 
that by adding more particles to the models 
the noise is reduced, then the apparition of a 
bar or spirals is delayed \citep{2003ApJ...587..638S}. 
However, all models have similar behavior during all time which
means that the general behavior of the models is more affected
by the spiral and/or bar activity than the noise as 
exposed by \citet{2003ApJ...587..638S}.

Finally, we made an orbital analysis in the bar reference frame
for those models where the bar was formed.
We proposed a very simple geometrically classification, 
in which we classified three types of orbits as follows:
compacts, along the bar, and
orbits trapped in the Lagrangian points $L_4$ and $L_5$.
Our main outcome was that after the bar formation, 
the compact like orbits increases their number
to reach around 5\% of the disk mass; the loop like orbits
increases their number to achieved around
10\% of the disk mass for all of these models; and 
the bar like orbits increases their number to attain around of
15\%, 13\%, 12\% and 16\% for fiducial,  s37\_z10MX,
s37\_z10MS, and s37\_z10MXS models, respectively. 
{\it Thus, we have found that these three geometrically
classified type of orbits, which are the ones
trapped in the Lagrangian points $L_3$, $L_4$, 
and $L_5$, encompass approximately
one-third of the disk total mass for these barred models.
Furthermore, a particle can change its orbit morphology
during its evolution.}

\section*{Acknowledgments}

%We would like to thank an anonymous referee for a constructive and helpful report,
%which has significantly improved this paper.

The authors gratefully acknowledge the very constructive comments offered by the
referee, Jerry Sellwood, which improved the presentation of this paper.
The authors also acknowledge support from the Mexican foundation CONACyT for
research grants.


\begin{thebibliography}{}

\bibitem[\protect\citeauthoryear{Athanassoula \& Sellwood}{1986}]{1986MNRAS.221..213A} Athanassoula, E., \& Sellwood, J.~A.\ 1986, MNRAS, 221, 213 

\bibitem[\protect\citeauthoryear{Athanassoula}{2003}]{2003LNP...626..313A} Athanassoula, L.\ 2003, Galaxies and Chaos, 626, 313 
 
\bibitem[\protect\citeauthoryear{Athanassoula}{2012}]{2012MNRAS.426L..46A} Athanassoula, E. 2012, MNRAS, 426, L46

\bibitem[\protect\citeauthoryear{Athanassoula}{2014}]{2014MNRAS.438L..81A} Athanassoula, E.\ 2014, MNRAS, 438, L81

\bibitem[\protect\citeauthoryear{Baba et al.}{2009}]{2009ApJ...706..471B} Baba, J., Asaki, Y., Makino, J., et al.\ 2009, ApJ, 706, 471 

\bibitem[\protect\citeauthoryear{Baba et al.}{2013}]{2013ApJ...763...46B} Baba, J., Saitoh, T.~R., \& Wada, K.\ 2013, ApJ, 763, 46

\bibitem[\protect\citeauthoryear{Bertin \& Lin}{1996}]{1996ssgd.book.....B} Bertin, G., \& Lin, C.~C.\ 1996, Spiral Structure in Galaxies: a Density Wave Theory (Cambridge MA, MIT Press)

\bibitem[\protect\citeauthoryear{Binney \& Tremaine}{1987}]{1987gady.book.....B} Binney, J., \& Tremaine, S. 1987, Galactic Dynamics (Princeton NJ, Princeton University Press)

\bibitem[\protect\citeauthoryear{Bottema}{2003}]{2003MNRAS.344..358B} Bottema, R.\ 2003, MNRAS, 344, 358

\bibitem[\protect\citeauthoryear{Buta et al.}{2009}]{butaetal2009} Buta, R. J., Knapen, J. H., Elmegreen, B. G. et al. 2009, AJ, 137, 4487

\bibitem[\protect\citeauthoryear{Carlberg \& Freedman}{1985}]{1985ApJ...298..486C} Carlberg, R.~G., \& Freedman, W.~L.\ 1985, ApJ, 298, 486

\bibitem[\protect\citeauthoryear{Chatzopoulos et al.}{2011}]{2011MNRAS.416..479C} Chatzopoulos, S., Patsis, P.~A., \& Boily, C.~M.\ 2011, MNRAS, 416, 479 

\bibitem[\protect\citeauthoryear{Dehnen}{2000}]{2000AGM....17..C01D} Dehnen, W.\ 2000, Astronomische Gesellschaft Meeting Abstracts, 17, 1 

\bibitem[\protect\citeauthoryear{Dehnen}{2002}]{2002JCoPh.179...27D} Dehnen, W.\ 2002, Journal of Computational Physics, 179, 27 

\bibitem[\protect\citeauthoryear{Dobbs \& Baba}{2014}]{dobbsbaba2014} Dobbs, C. \& Baba, J. 2014, PASA, 31, 35

\bibitem[\protect\citeauthoryear{D'Onghia et al.}{2010}]{2010ApJ...709.1138D} D'Onghia, E., Springel, V., Hernquist, L., \& Keres, D.\ 2010, ApJ, 709, 1138 

\bibitem[\protect\citeauthoryear{D'Onghia et al.}{2013}]{2013ApJ...766...34D} D'Onghia, E., Vogelsberger, M., \& Hernquist, L.\ 2013, ApJ, 766, 34 

\bibitem[\protect\citeauthoryear{Foyle et al.}{2011}]{2011ApJ...735..101F} Foyle, K., Rix, H. W., Dobbs, C. L., Leroy, A. K., \& Walter, F. 2011, ApJ, 735, 101 

\bibitem[\protect\citeauthoryear{Fuchs et al.}{2005}]{2005A&A...444....1F} Fuchs, B., Dettbarn, C., \& Tsuchiya, T.\ 2005, A\&A, 444, 1 

\bibitem[\protect\citeauthoryear{Fujii et al.}{2011}]{2011ApJ...730..109F} Fujii, M.~S., Baba, J., Saitoh, T.~R., et al.\ 2011, ApJ, 730, 109 

\bibitem[\protect\citeauthoryear{Gauthier et al.}{2006}]{2006ApJ...653.1180G} Gauthier, J.-R., Dubinski, J., \& Widrow, L.~M.\ 2006, ApJ, 653, 1180 

\bibitem[\protect\citeauthoryear{Gerin et al.}{1990}]{1990A&A...230...37G} Gerin, M., Combes, F., \& Athanassoula, E.\ 1990, A\&A, 230, 37 

\bibitem[\protect\citeauthoryear{Goldreich \& Lynden-Bell}{1965}]{1965MNRAS.130..125G} Goldreich, P., \& Lynden-Bell, D.\ 1965, MNRAS, 130, 125 

\bibitem[\protect\citeauthoryear{Grand et al.}{2012a}]{2012MNRAS.421.1529G} Grand, R. J. J., Kawata, D., \& Cropper, M.\ 2012a, MNRAS, 421, 1529

\bibitem[\protect\citeauthoryear{Grand et al.}{2012b}]{2012MNRAS.426..167G} Grand, R. J. J., Kawata, D., \& Cropper, M.\ 2012b, MNRAS, 426, 167

%\bibitem[\protect\citeauthoryear{Holley-Bockelmann et al.}{2005}]{2005MNRAS.363..991H} Holley-Bockelmann, K., Weinberg, M., \& Katz, N.\ 2005, MNRAS, 363, 991 

\bibitem[\protect\citeauthoryear{Julian \& Toomre}{1966}]{1966ApJ...146..810J} Julian, W.~H., \& Toomre, A.\ 1966, ApJ, 146, 810 

\bibitem[\protect\citeauthoryear{Klypin et al.}{2009}]{2009MNRAS.398.1027K} Klypin, A., Valenzuela, O., Col{\'{\i}}n, P., \& Quinn, T.\ 2009, MNRAS, 398, 1027 

\bibitem[\protect\citeauthoryear{Kuijken \& Dubinski}{1994}]{1994MNRAS.269...13K} Kuijken, K., \& Dubinski, J.\ 1994, MNRAS, 269, 13 

\bibitem[\protect\citeauthoryear{Kuijken \& Dubinski}{1995}]{1995MNRAS.277.1341K} Kuijken, K., \& Dubinski, J.\ 1995, MNRAS, 277, 1341 

\bibitem[\protect\citeauthoryear{Lin \& Shu}{1964}]{1964ApJ...140..646L} Lin, C.~C., \& Shu, F.~H.\ 1964, ApJ, 140, 646

\bibitem[\protect\citeauthoryear{Mart{\'{\i}}nez-Garc{\'{\i}}a \& Gonz{\'a}lez-L{\'o}pezlira}{2013}]{2013ApJ...765..105M} Mart{\'{\i}}nez-Garc{\'{\i}}a, E.~E., \& Gonz{\'a}lez-L{\'o}pezlira, R.~A.\ 2013, ApJ, 765, 105 

\bibitem[\protect\citeauthoryear{Masset \& Tagger}{1997}]{1997A&A...322..442M} Masset, F., \& Tagger, M.\ 1997, A\&A, 322, 442 

\bibitem[\protect\citeauthoryear{Mata-Ch{\'a}vez et al.}{2014}]{2014arXiv1408.3670M} Mata-Ch{\'a}vez, D., G{\'o}mez, G.~C., \& Puerari, I.\ 2014, MNRAS, 444, 3756

\bibitem[\protect\citeauthoryear{Meidt et al.}{2009}]{2009ApJ...702..277M} Meidt, S. E., Rand, R.~J., \& Merrifield, M.~R.\ 2009, ApJ, 702, 277 

\bibitem[\protect\citeauthoryear{Merritt \& Sellwood}{1994}]{1994ApJ...425..551M} Merritt, D., \& Sellwood, J.~A.\ 1994, ApJ, 425, 551 

\bibitem[\protect\citeauthoryear{Merrifield et al.}{2006}]{2006MNRAS.366L..17M} Merrifield, M. R., Rand, R. J., \& Meidt, S.~E.\ 2006, MNRAS, 366, L17 

\bibitem[\protect\citeauthoryear{Minchev \& Famaey}{2010}]{apj_722_1_112} Minchev, I., \& Famaey, B. 2010, ApJ, 722, 112

\bibitem[\protect\citeauthoryear{Okamoto et al.}{2015}]{2015PASJ...67...63O} Okamoto, T., Isoe, M., \& Habe, A.\ 2015, PASJ, 67, 63 

\bibitem[\protect\citeauthoryear{Puerari \& Dottori}{1992}]{1992A&AS...93..469P} Puerari, I., \& Dottori, H.~A.\ 1992, A\&AS, 93, 469 

\bibitem[\protect\citeauthoryear{Puerari et al.}{2000}]{puerarietal2000} Puerari, I., Block, D. L., Elmegreen, B. G., Frogel, J. A., Eskridge, P. B. 2000, A\&A, 359, 932

\bibitem[\protect\citeauthoryear{Roca-F{\`a}brega et al.}{2013}]{2013MNRAS.432.2878R} Roca-F{\`a}brega, S., Valenzuela, O., Figueras, F., et al.\ 2013, MNRAS, 432, 2878 

\bibitem[\protect\citeauthoryear{Romeo}{1998}]{1998AA...335..922R} Romeo, A. 1998, A\&A, 335, 922

\bibitem[\protect\citeauthoryear{Saha \& Elmegreen}{2016}]{sahaelmegreen2016} Saha, K. \& Elmegreen, B. G. 2016, ApJ 826, L21

\bibitem[\protect\citeauthoryear{Sellwood \& Carlberg}{1984}]{1984ApJ...282...61S} Sellwood, J.~A., \& Carlberg, R.~G.\ 1984, ApJ, 282, 61 

\bibitem[\protect\citeauthoryear{Sellwood}{2000}]{2000Ap&SS.272...31S} Sellwood, J.~A.\ 2000, Ap\&SS, 272, 31 

\bibitem[\protect\citeauthoryear{Sellwood}{2003}]{2003ApJ...587..638S} Sellwood, J.~A.\ 2003, ApJ, 587, 638 

\bibitem[\protect\citeauthoryear{Sellwood}{2010}]{2010arXiv1001.5430S} Sellwood, J.~A.\ 2010, arXiv:1001.5430 

\bibitem[\protect\citeauthoryear{Sellwood}{2011}]{2011MNRAS.410.1637S} Sellwood, J.~A.\ 2011, MNRAS, 410, 1637 

\bibitem[\protect\citeauthoryear{Sellwood}{2013}]{2013ApJ...769L..24S} Sellwood, J.~A.\ 2013, ApJ, 769, L24 

\bibitem[\protect\citeauthoryear{Sellwood \& Binney}{2002}]{2002MNRAS.336..785S} Sellwood, J.~A., \& Binney, J.~J.\ 2002, MNRAS, 336, 785 

\bibitem[\protect\citeauthoryear{Sellwood}{2013}]{2013pss5.book..923S} Sellwood, J.~A.\ 2013, Planets, Stars and Stellar Systems.~Volume 5: Galactic Structure and Stellar Populations, 923

\bibitem[\protect\citeauthoryear{Sellwood \& Carlberg}{2014}]{2014ApJ...785..137S} Sellwood, J.~A., \& Carlberg, R.~G.\ 2014, ApJ, 785, 137 

\bibitem[\protect\citeauthoryear{Speights \& Westpfahl}{2012}]{2012ApJ...752...52S} Speights, J. C., \& Westpfahl, D. J.\ 2012, ApJ, 752, 52 

\bibitem[\protect\citeauthoryear{Tremaine \& Weinberg}{1984}]{tremaine1984} Tremaine, S., \& Weinberg, M. D.\ 1984, MNRAS, 209, 729

\bibitem[\protect\citeauthoryear{Teuben}{1995}]{1995ASPC...77..398T} Teuben, P.\ 1995, Astronomical Data Analysis Software and Systems IV, 77, 398 

\bibitem[\protect\citeauthoryear{Toomre}{1981}]{1981seng.proc..111T} Toomre, A.\ 1981, in Structure and Evolution of Normal Galaxies (Cambridge University Press), 111 

\bibitem[\protect\citeauthoryear{Valencia-Enr{\'{\i}}quez \& Puerari}{2014}]{2014ASPC..480..145V} Valencia-Enr{\'{\i}}quez, D., \& Puerari, I.\ 2014, ASP Conference Series, 480, 145 

\bibitem[\protect\citeauthoryear{Wada et al.}{2011}]{2011ApJ...735....1W} Wada, K., Baba, J., \& Saitoh, T.~R.\ 2011, ApJ, 735, 1 

\bibitem[\protect\citeauthoryear{Weinberg \& Katz}{2007}]{2007MNRAS.375..425W} Weinberg, M.~D., \& Katz, N.\ 2007, MNRAS, 375, 425 

\bibitem[\protect\citeauthoryear{Weinberg \& Katz}{2007}]{2007MNRAS.375..460W} Weinberg, M.~D., \& Katz, N.\ 2007, MNRAS, 375, 460

\end{thebibliography}
\end{document}